\documentclass[aps,prX,notitlepage,superscriptaddress,amssymb,graphicx]{revtex4-1}

\usepackage{tocvsec2}
\usepackage{etoolbox}

\usepackage{graphicx}
\usepackage{dcolumn} 
\usepackage{nicefrac}
\usepackage{mathrsfs}
\newcommand{\diff}{\mathrm{d}}
\usepackage{suffix}
\usepackage{mathtools}
\def\be{\begin{equation}}
\def\ee{\end{equation}}
\newcommand{\fref}[1]{Fig.~\ref{fig:#1}}
\newcommand{\eref}[1]{Eq.~(\ref{eq:#1})}

\DeclarePairedDelimiterX\MeijerM[3]{\lparen}{\rparen}%
{\begin{smallmatrix}#1 \\ #2\end{smallmatrix}\delimsize\vert\,#3}

\newcommand\MeijerG[8][]{%
  G^{\,#2,#3}_{#4,#5}\MeijerM[#1]{#6}{#7}{#8}}

\WithSuffix\newcommand\MeijerG*[7]{%
  G^{\,#1,#2}_{#3,#4}\MeijerM*{#5}{#6}{#7}}
\newcommand{\Py}{Ni$_{81}$Fe$_{19}$}
\patchcmd{\thebibliography}{\section*{\refname}}{}{}{}

\begin{document}
\title{Nanometre-scale probing of spin waves using single electron spins}

\author{T. van der Sar$^*$}
\affiliation{Department of Physics, Harvard University, 17 Oxford St., Cambridge, MA 02138, USA.}
\altaffiliation{These authors contributed equally to this work}

\author{F. Casola$^*$}
\affiliation{Harvard-Smithsonian Center for Astrophysics, 60 Garden St., Cambridge, MA 02138, USA.}
\affiliation{Department of Physics, Harvard University, 17 Oxford St., Cambridge, MA 02138, USA.}

\author{R. Walsworth}
\affiliation{Harvard-Smithsonian Center for Astrophysics, 60 Garden St., Cambridge, MA 02138, USA.}
\affiliation{Department of Physics, Harvard University, 17 Oxford St., Cambridge, MA 02138, USA.}

\author{A. Yacoby}
\affiliation{Department of Physics, Harvard University, 17 Oxford St., Cambridge, MA 02138, USA.}

\date{\today}

\maketitle

\textbf{Correlated-electron systems support a wealth of magnetic excitations, ranging from conventional spin waves to exotic fractional excitations in low-dimensional or geometrically-frustrated spin systems\cite{Schollwock04,Lacroix11}. Probing such excitations on nanometre length scales is essential for unravelling the underlying physics and developing new spintronic nanodevices\cite{Chappert07,Parkin08,Sinova12,Fert13}. However, no established technique provides real-space, few-nanometre-scale probing of correlated-electron magnetic excitations under ambient conditions. Here we present a solution to this problem using magnetometry based on single nitrogen-vacancy (NV) centres in diamond\cite{Degen08,Taylor08}. We focus on spin-wave excitations in a ferromagnetic microdisc, and demonstrate local, quantitative, and phase-sensitive detection of the spin-wave magnetic field at $\sim 50$ nm from the disc. We map the magnetic-field dependence of spin-wave excitations by detecting the associated local reduction in the disc's longitudinal magnetization. In addition, we characterize the spin-noise spectrum by NV-spin relaxometry\cite{Steinert13,Tetienne13}, finding excellent agreement with a general analytical description of the stray fields produced by spin-spin correlations in a 2D magnetic system. These complementary measurement modalities pave the way towards imaging the local excitations of systems such as ferromagnets and antiferromagnets, skyrmions\cite{Nagaosa13}, atomically assembled quantum magnets\cite{Spinelli14}, and spin ice\cite{Lacroix11}.} \\

\noindent
Pushing the frontiers of condensed-matter magnetism requires tools to probe magnetic excitations with nanometer-scale spatial resolution. Despite recent progress with real-space techniques\cite{Spinelli14,Lee10,Vansteenkiste09,Pigeau10,Demokritov08,Vasyukov13,Nowack13}, a wide range of interesting magnetic phenomena in correlated-electron materials remains experimentally inaccessible because of the required combination of resolution, magnetic-field sensitivity, and environmental compatibility. Here, we show that a magnetic imaging system employing single nitrogen-vacancy (NV) centres in diamond can overcome this challenge.\\

\noindent
The S=1 electronic spin of the NV centre in diamond is an atom-sized magnetic field sensor that can be brought within a few nanometres of a sample and readily interrogated with optically detected magnetic resonance\cite{Balasubramanian2008}. NV-centre magnetometry\cite{Degen08,Taylor08} has provided unprecedented room-temperature magnetic imaging with nanometre-scale resolution\cite{Balasubramanian2008,Grinolds13,Mamin13,Tetienne14} and single-proton-spin sensitivity\cite{Sushkov14}, and has been used to study nanoscale bio-magnetism\cite{Sage13,SchaferNolte14}. However, NV centres have only recently emerged as probes of collective spin dynamics in correlated-electron systems\cite{Tetienne14,Wolfe14}. In this Letter, we demonstrate that single-NV magnetic imaging is a powerful tool for nanometre-scale, quantitative, and non-perturbative detection of spin-wave excitations. We present complementary measurement techniques to study spin-wave excitations over a broad range of magnetic fields and frequencies, as well as a method to characterize spin-spin correlations. These methods may be directly applied to open problems of current interest, such as real-space imaging of skyrmion core dynamics\cite{Nagaosa13} or imaging spin-wave excitations in atomically-assembled magnets\cite{Spinelli14} as a function of temperature.\\

\noindent
As a model system, we consider a ferromagnetic microdisc (\Py) fabricated on top of a diamond chip that contains NV centres implanted at $\sim$50 nm below the surface (Fig. 1a-b). We use an on-chip coplanar waveguide to generate microwave magnetic fields to control the NV spin state and to drive spin-wave excitations in the disc. We optically address individual NV centres using a scanning confocal microscope (Fig. 1b) and read out the NV spin state through spin-dependent photoluminescence (Supplementary Section I).\\ 

\begin{figure}[h!]
\centering
\includegraphics[width=\textwidth]{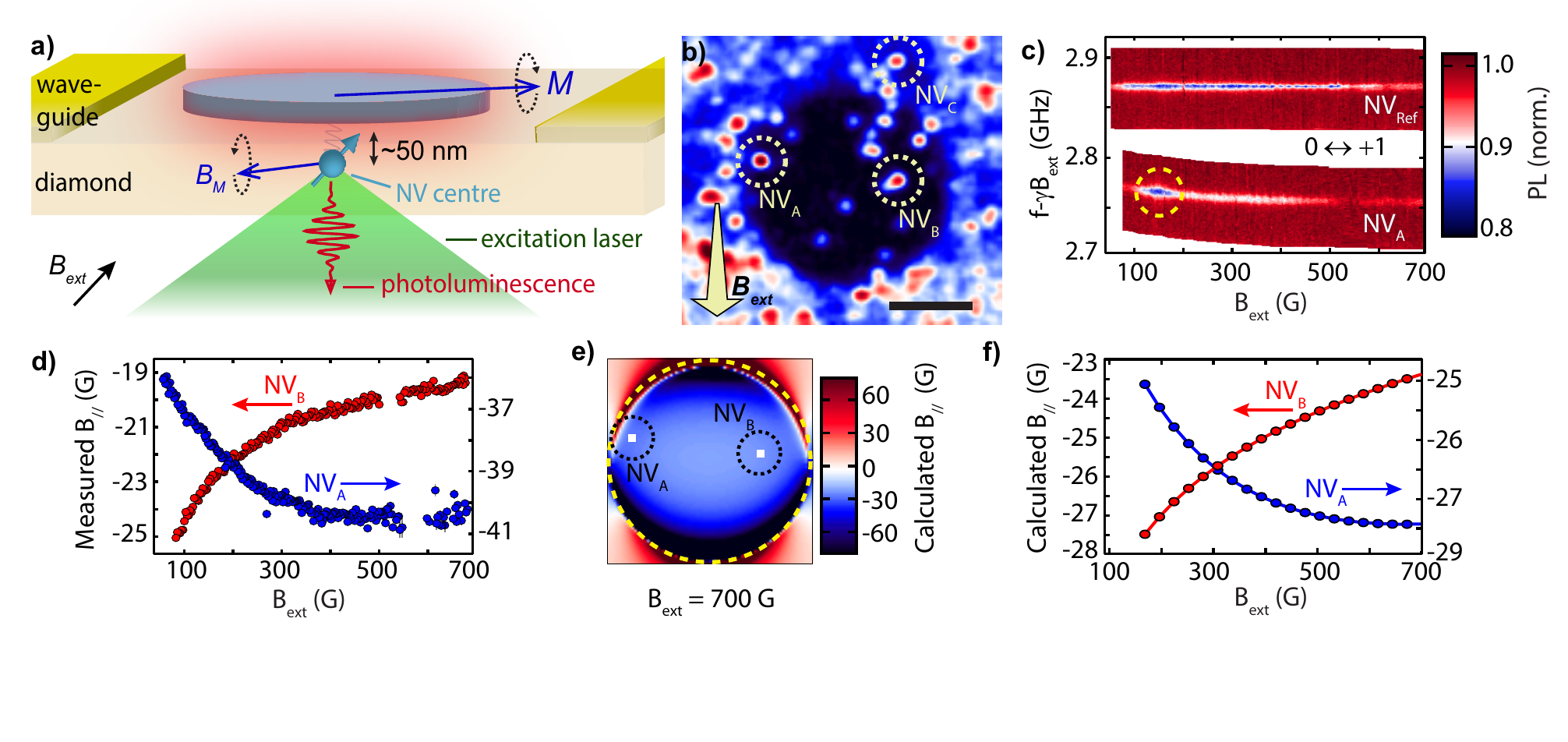}
\vspace{-1.5cm}
\caption{ \textbf{Nanoscale probing of the static and dynamic magnetic fields generated by a ferromagnetic microdisc.} \textbf{a,} , As a model system to study magnetic excitations, we consider a ferromagnetic microdisc (\Py) fabricated on top of a diamond surface. NV centres implanted at $\sim 50$ nm below the surface sense the local magnetic fields $B_M$ generated by the magnetization $M$. \textbf{b,} Scanning confocal microscopy image showing a photoluminescence map of NV centres close to the disc. Scale bar: 3 $\mu$m. The external static magnetic field $B_{\mathrm{ext}}$ is applied along the axis of target NV centres (see methods) \textbf{c,} Optically detected electron spin resonance (ESR) traces of the $m_s = 0 \leftrightarrow +1$ transition of NV$_{\mathrm{A}}$ (close to the disc) and NV$_{\mathrm{ref}}$ (at $\sim 11$ $\mu$m from the disc centre). From the ESR frequency of NV$_{\mathrm{ref}}$ we extract the external magnetic field $B_{\mathrm{ext}}$ (Supplementary Section II). From the difference between NV$_{\mathrm{A}}$ and NV$_{\mathrm{ref}}$ we extract the disc stray field at the site of NV$_{\mathrm{A}}$. The dashed circle indicates power broadening caused by amplification of the drive field by a spin-wave excitation, as discussed in Fig.\ 2. 
\textbf{d,} Projection onto the NV axis of the measured disc stray field as a function of external field, showing opposite behaviour at the sites of NV$_{\mathrm{A}}$ and NV$_{\mathrm{B}}$. The sign of the field is relative to $B_{\mathrm{ext}}$. \textbf{e,} Calculated spatial profile of the projection of the disc stray field onto the NV axis in a plane 50 nm below the disc. $B_{\mathrm{ext}}$ = 700 G. \textbf{f,} Numerically calculated projection of the disc stray field onto the NV axis as a function of the external field at the sites of NV$_{\mathrm{A}}$ and NV$_{\mathrm{B}}$, in qualitative good agreement with the measurements in \textbf{d}.}
\label{fig:fig1}
\end{figure}

\noindent
Characterization of the static magnetization forms the basis for understanding the excitations of a magnetic system. Using individual NV centres close to the disc, we locally characterize the magnetization as a function of an externally applied static magnetic field $B_{\textrm{ext}}$. We measure the electron spin resonance (ESR) frequency of an NV centre close to the disc (NV$_{\mathrm{A}}$ in Fig. 1b) and a reference NV centre (NV$_{\mathrm{ref}}$) far from the disc (Fig. 1c). By comparing these ESR frequencies and knowing the NV gyromagnetic ratio $\gamma = $2.8 MHz/G, we determine the stray magnetic field of the disc at the location of NV$_{\mathrm{A}}$ (see Methods). Fig 1d shows the projection of this ``disc stray field" onto the NV axis, $B_{//}$, as a function of $B_{\textrm{ext}}$.\\

\noindent
The local nature of the disc's magnetization becomes clear by comparing the measured disc stray field at two locations (NV$_{\mathrm{A}}$ and NV$_{\mathrm{B}}$, Fig.1b). At both NV$_{\mathrm{A}}$ and NV$_{\mathrm{B}}$, this field opposes the external field (Fig. 1d), as expected from a calculated spatial field profile (Fig. 1e). However, as $B_{\textrm{ext}}$ is decreased, the change in the disc stray field at NV$_{\mathrm{A}}$ is remarkably opposite to that at NV$_{\mathrm{B}}$ (Fig 1d). This behaviour is qualitatively in good agreement with numerical calculations of the disc's magnetization and the associated disc stray field as a function of $B_{\textrm{ext}}$ (Fig. 1f). These calculations indicate that as $B_{\textrm{ext}}$ is decreased, the magnetization becomes less homogeneous, with spins at the disc's edge reorienting first. The opposite behaviour of the disc stray field at NV$_{\mathrm{A}}$ and NV$_{\mathrm{B}}$ is a direct result of the differently varying local magnetization (Supplementary Section VI), and would not be observed in a far-field measurement.\\

\noindent
Spin-wave excitations consist of collectively precessing spins in a magnetically ordered system. It was recently proposed\cite{Trifunovic14} that detection of the time-varying stray magnetic fields generated by spin-wave excitations in small ferromagnets may be exploited to strongly amplify the sensitivity of single NV-centre magnetometry. Here we employ a 'resonant' detection technique to locally sense the spin-wave stray magnetic field, demonstrating the first single-spin detection of on-chip magnetic-field amplification by a ferromagnet. We apply a microwave (MW) magnetic field to drive spin-wave excitations in the disc, choosing the MW frequency such that it is resonant with the ESR frequency of a target NV centre. The spins in the disc respond and generate a magnetic field $\mathbf{B}^i_{M}$ which interferes with the drive field $\mathbf{B}_{D}$. Transformed into a frame rotating at the ESR frequency, these fields sum (inset Fig. 2b) to give the total field $|\mathbf{b}^i|$ = $|\mathbf{b}^i_M + \mathbf{b}_D|$ = $f_R^{NV_i}/ \gamma$ driving spin rotations (Rabi oscillations) of NV$_i$ at a rate $f_R^{NV_i}$. As we tune the NV ESR frequency using B$_{\textrm{ext}}$ we observe a striking difference between the Rabi frequency of nearby NV centres and the Rabi frequency of a far-away, reference NV$_{\mathrm{ref}}$ (Fig. 2a). This difference becomes even clearer by plotting the ratio $\frac{f_R^{NV_i}}{f_R^{NV_{\mathrm{ref}}}}$, which corrects for any frequency-dependent delivery of microwaves (MWs) through our setup (Fig. 2b).\\

\begin{figure}[h!]
\centering
\includegraphics[width=\textwidth]{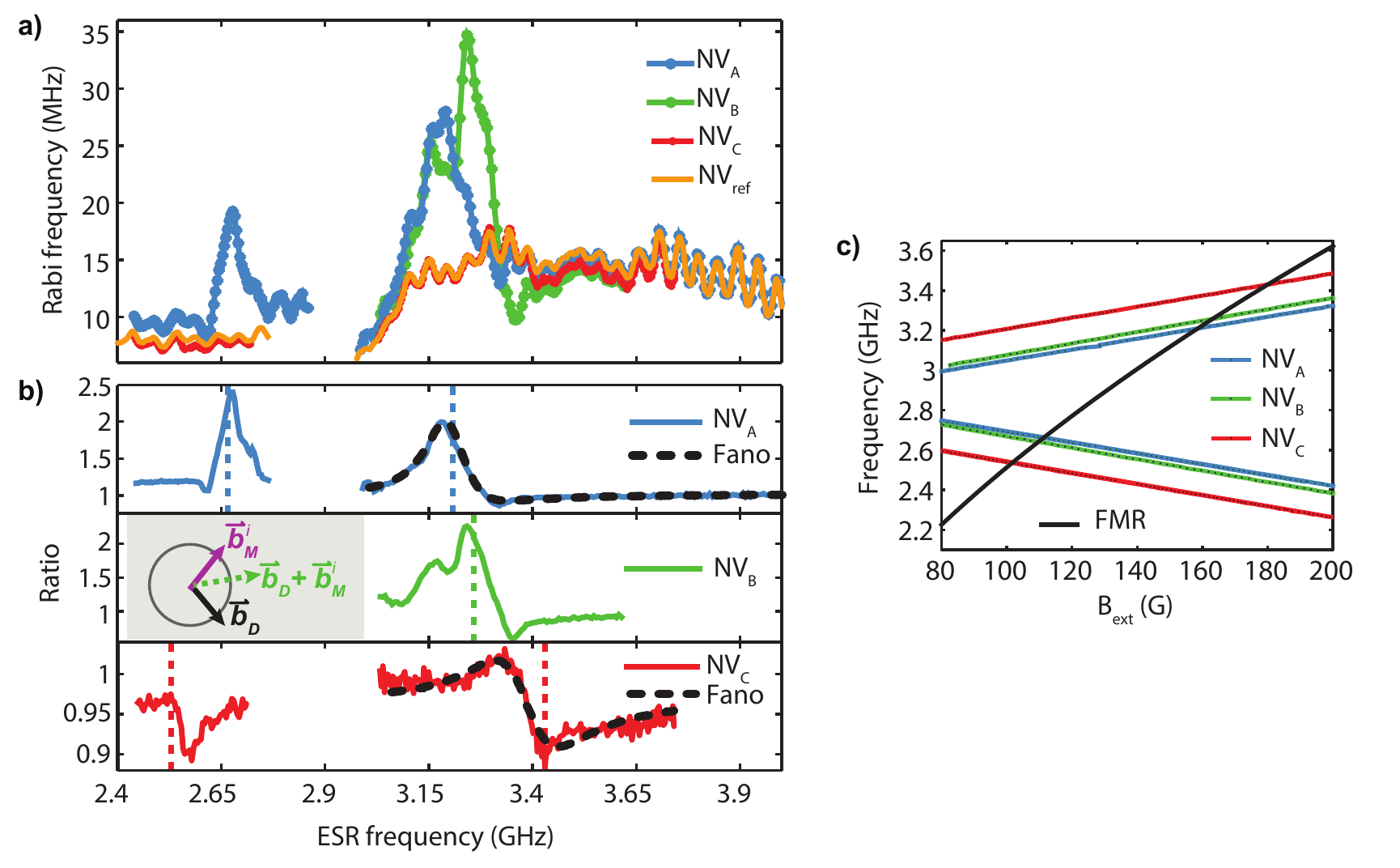}
\caption{ \textbf{Resonant detection of driven spin-wave excitations.} \textbf{a,} Measured Rabi frequency as a function of ESR frequency for three NV centres close to the disc (NV$_{\mathrm{A}}$, NV$_{\mathrm{B}}$, and NV$_{\mathrm{C}}$, see Fig. 1b) and reference NV centre NV$_{\mathrm{ref}}$ far from the disc. We tune the NV ESR frequency using $B_{\mathrm{ext}}$ and drive Rabi oscillations by applying a MW magnetic field at the ESR frequency. The ESR frequency range below (above) 2.87 GHz corresponds to the $m_s = 0 \leftrightarrow -1 (m_s = 0  \leftrightarrow +1)$ transition. \textbf{b,} Ratio of the Rabi frequency of NV$_i$ ($i=$A, B, C) over the Rabi frequency of NV$_{\mathrm{ref}}$, as measured in Fig. 2a. Pronounced resonances are visible where the NV centre ESR frequency matches the numerically calculated ferromagnetic resonance (FMR) of the disc, which occurs at the vertical dashed lines (see Fig. 2c). The grey inset depicts the interference between the AC magnetic field generated by a spin-wave excitation $\mathbf{b}_M^i$  at the site of NV$_i$ ($i=$A, B, C) and the driving field $\mathbf{b}_D$ in a frame rotating at the NV centre ESR frequency. \textbf{c,} Measured ESR frequencies and numerically calculated FMR frequency as a function of magnetic field. }
\label{fig:fig2}
\end{figure}

\noindent
Numerical calculations of the spin-wave spectrum of the disc (Supplementary Section VII) indicate that the resonances in Fig. 2b occur when the NV centre ESR frequency matches the frequency of the lowest order spin-wave resonance of the disc (Fig 2c). This mode - the ferromagnetic resonance (FMR) - is efficiently excited because the driving field is spatially uniform (Supplementary Section VII). The observed resonance is described by:
\begin{equation}
\frac{f_R^{NV_i}}{f_R^{NV_{\mathrm{ref}}}}(f) \propto \sqrt{1 + r(f)^2 + 2 r(f)\cos[\theta(f)]},
\label{eq:Fano}
\end{equation}
where $r(f) = \frac{|\mathbf{b}^i_M|}{|\mathbf{b}_D|}(f)$, and $\theta(f)$ is the angle between $\mathbf{b}^i_M$ and $\mathbf{b}_D$, determined by the dynamic susceptibility of the ferromagnet and the location of the NV centre. To illustrate the validity of this model, we fit the resonances of NV$_{\mathrm{A}}$ and NV$_{\mathrm{C}}$ using eq. 1, assuming a simple damped oscillator response for $r(f)$ and $\theta(f)$ (Supplementary Section III). The resulting Fano-lineshape accurately describes the observed interference, demonstrating that this technique is sensitive to both the amplitude and phase of the spin-wave magnetic field. Possible deviations from this model, such as the double peak structure of NV$_{\mathrm{B}}$, may result from frequency-dependence of the spatial profile of the spin-wave excitation or fabrication-related imperfections. The amplification of the MW field also explains the power broadening of the ESR spectra at low applied magnetic fields $B_{\mathrm{ext}}$, as observed in Fig. 1c.\\

\noindent
Characterizing the magnetic excitation spectrum in a correlated-electron system, as well as addressing other problems of interest such as imaging magnetic-vortex\cite{Vansteenkiste09,Pigeau10} or skyrmion core dynamics\cite{Nagaosa13}, requires a detection scheme that operates over a broad frequency range. To this end, we developed an 'off-resonant' detection technique that can detect a sample's spin dynamics even when the NV centre ESR frequency is far detuned from the frequency of these dynamics. The idea is to drive a spin-wave excitation in the sample with a microwave magnetic field and detect the resulting change in the stray magnetic field by applying a multi-pulse sensing sequence to the NV centre\cite{Lange11}. As an experimental demonstration for the ferromagnetic microdisc, we use the sequence in Fig. 3a, which decouples the NV spin from magnetic fields that are static on the time scale of the sequence, while being sensitive to changes in the local stray magnetic field induced by the MW driving. We relate the phase $\phi$ accumulated by the NV centre at the end of the sequence to an effective magnetic field $B_{\textrm{eff}} = \phi/(\gamma T) $ oriented along the NV-axis and averaged over the duration $T$ of the MW driving.\\
\begin{figure}[h!]
\centering
\includegraphics[width=\textwidth]{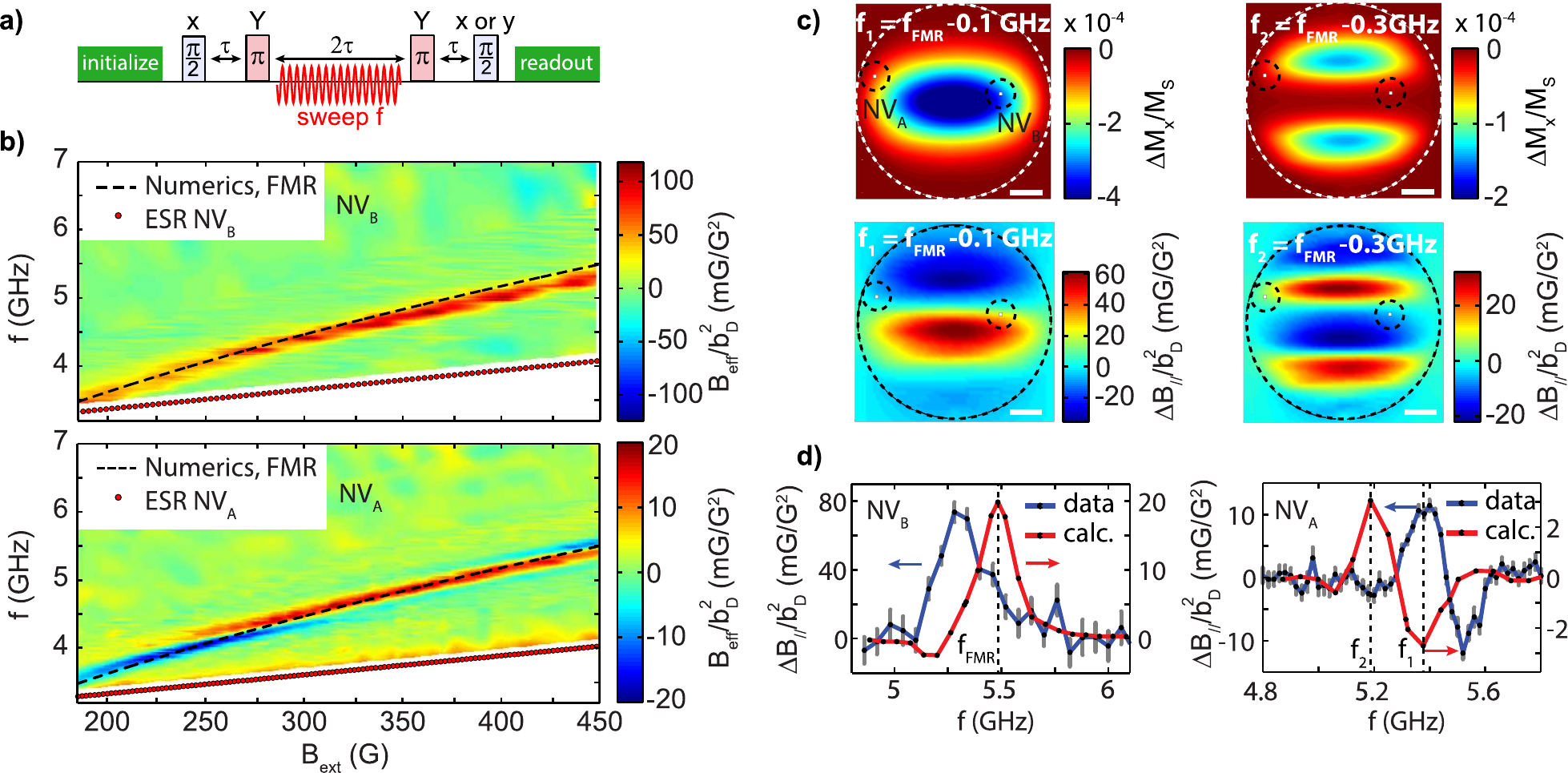}
\caption{ \textbf{Non-resonant, field-dependent detection of driven spin-wave excitations.} \textbf{a,} Measurement sequence. The first $\pi/2$ pulse prepares an NV spin superposition, which is input into an echo sequence with two  $\pi$-pulses. Synchronized with this sequence, we apply microwave (MW) driving at frequency $f$ during the central $2 \tau$ period of free evolution to excite spin-wave excitations in the disc. We read out the final phase $\phi$ of the NV spin state by measuring the projection on the x- and y-axis. \textbf{b,} Experimentally-determined effective magnetic field $B_{\textrm{eff}}$ at NV$_{\mathrm{B}}$ (top panel) and NV$_{\mathrm{A}}$ (lower panel) as a function of MW driving frequency $f$ and external static magnetic field $B_{\mathrm{ext}}$.  $B_{\textrm{eff}}$ is normalized by the square of the drive field $|\mathbf{b}_D |^2$ measured on-chip using NV$_{\mathrm{ref}}$  (see Methods and Supplementary Section IV). The dashed line is a numerical calculation of the ferromagnetic resonance (FMR) of the disc. For NV$_{\mathrm{A}}$, the AC Stark effect is visible just above the ESR frequency over the entire magnetic field range. We use the sign of the Stark effect to determine the sign of $B_{\textrm{eff}}$ in these measurements (Supplementary Section V). Scale bar: 1 $\mu$m. \textbf{c,} Top panel: numerically calculated spatial profile of the time-averaged change in the disc's longitudinal magnetization $\Delta M_x$ relative to the saturation magnetization $M_S$ under spatially uniform driving with a 5 G MW field. Bottom panel: associated stray magnetic field $\Delta B_{//}$ in the NV-plane for two MW frequencies $f_1$ and $f_2$ close to the FMR. We use an external static field $B_{\mathrm{ext}}$ = 450 G, corresponding to the highest field used in the measurements in Fig. 4b, at which we expect the disc magnetization to be the most homogeneous and resemblant of the calculated magnetization. \textbf{d,} Comparison of measured and calculated FMR lineshapes at $B_{\mathrm{ext}}$ = 450 G. The sign, amplitude, and width of the lineshapes accurately match the calculations. The 4\% difference in frequency presumably results from a difference in the disc's saturation magnetization and/or fabrication-related imperfections.}
\label{fig:fig3}
\end{figure}

\noindent
For both NV$_{\mathrm{A}}$ and NV$_{\mathrm{B}}$, we observe a clear resonance which agrees well with numerical calculations of the FMR frequency of the disc (Fig. 3b, Supplementary Section VII). The resonance follows a Kittel-like law $f_{\textrm{FMR}} = \gamma \sqrt{B_{\textrm{ext}} \left( B_{\textrm{ext}} + A \right)}$, where $A$ is a free parameter, characteristic of spin-wave excitations in a thin ferromagnet with in-plane magnetization. We therefore conclude that the observed resonance corresponds to the FMR of the disc. Furthermore, we observe striking differences in the lineshape of the resonances detected with NV$_{\mathrm{A}}$ and NV$_{\mathrm{B}}$ (Fig. 3b). To gain more insight into the origin of these differences, we now discuss how spin-wave magnetic fields manifest themselves in these measurements.\\

\noindent
Firstly, we note that the off-resonant driving field may contribute to $B_{\textrm{eff}}$ through the dynamical (i.e., AC) Stark effect\cite{Wei97}, which describes a renormalization of the NV spin transition frequencies due to the driving field. However, this effect quickly diminishes for increasing detuning and we estimate it to play a minor role in Fig. 3b (Supplementary Section V). Secondly, the excitation of a spin-wave mode reduces the time-averaged magnetization of the disc within a spatial region set by the mode profile\cite{Lee10}. The corresponding time-averaged change $\Delta B_{//} (f)$ the disc field at the location of the NV centre affects $B_{\textrm{eff}}$. Because of the close proximity of the NV centres, $\Delta B_{//} (f)$ strongly depends on the NV-centre location with respect to the spin-wave mode profile (Fig. 3c). In addition, this profile depends on frequency (Fig. 3c), affecting the lineshape of $\Delta B_{//} (f)$. At $B_{\textrm{ext}} = 450$ G we find a remarkably good agreement of the sign, width, and shape of the measured FMR signal with calculations (Fig. 3d) given geometrical uncertainties related to the optical resolution ($\sim 400$ nm), disc fabrication, NV implantation depth, and oxidation. However, we note that these calculations and/or our model do not account for the change in FMR lineshape observed at NV$_{\mathrm{A}}$ as we decrease $B_{\textrm{ext}}$. Such strong sensitivity on location highlights the unique possibilities NV centres offer to study spin dynamics quantitatively and with nanometre-scale resolution.\\

\noindent
Spin noise contains valuable information about a system's magnetic excitation spectrum and is present even in the absence of driving. Here, we spectrally probe spin noise in the disc by measuring the spin relaxation rates of a proximal NV centre (NV$_{\mathrm{A}}$), which depend on the strength of the magnetic field generated by the spin noise at the NV centre ESR frequencies\cite{Steinert13,Tetienne13}. As we lower $B_{\textrm{ext}}$ and thereby change the NV ESR frequencies relative to the spin-noise spectrum, we first find the $m_s = 0 \leftrightarrow +1$ and then the $m_s = 0 \leftrightarrow -1$ relaxation rate to increase by over an order of magnitude (Fig. 4a-b), indicating a dramatic increase in the noise at the ESR frequencies.\\

\begin{figure}[h!]
\centering
\includegraphics[width=0.75\textwidth]{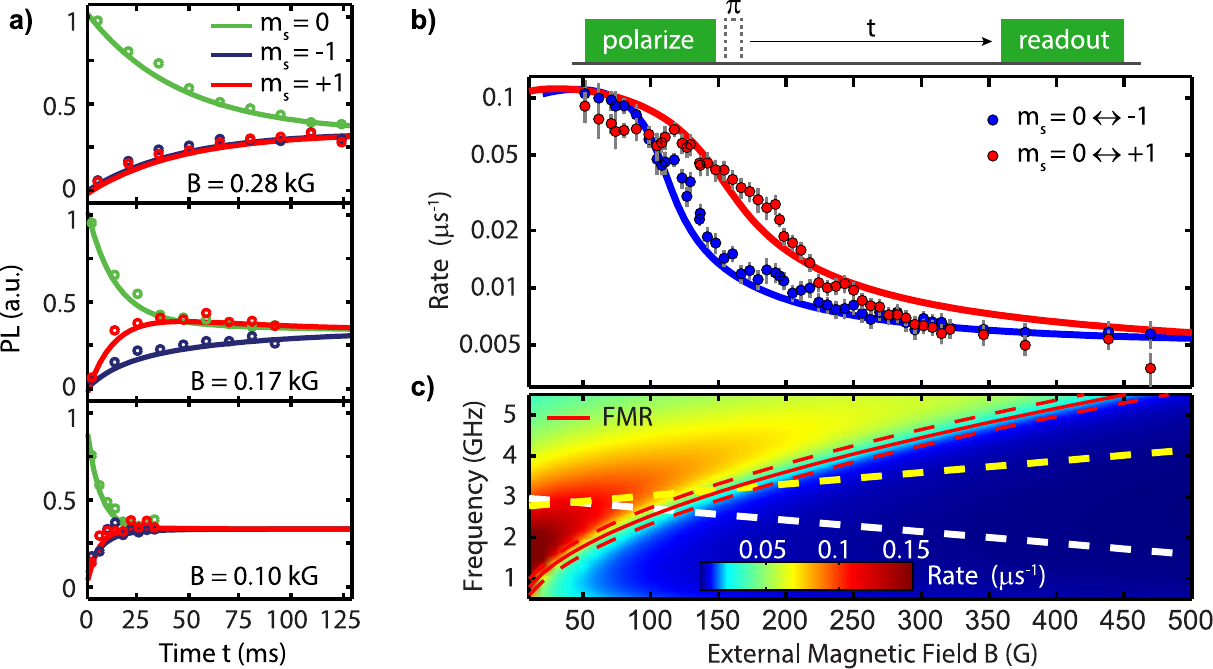}
\caption{ \textbf{Probing the spin-noise spectrum.} \textbf{a,} Spin relaxation measurements for NV$_{\mathrm{A}}$ at three different external static magnetic fields $B_{\mathrm{ext}}$. The NV centre is prepared in each of its 3 spin eigenstates ($m_s =$0, -1, +1), and the spin-dependent photoluminescence is measured as a function of waiting time $t$. We extract the $m_s = 0 \leftrightarrow +1$ and the $m_s = 0 \leftrightarrow -1$ relaxation rates by fitting with a three-level model (Supplementary Section VI). \textbf{b,} Measured NV spin relaxation rates as a function of $B_{\mathrm{ext}}$. As we lower $B_{\mathrm{ext}}$, the $m_s = 0 \leftrightarrow +1$ and the $m_s = 0 \leftrightarrow -1$ relaxation rates consecutively increase by about an order of magnitude. Dots are measured data. Solid lines are from calculations of the magnetic noise spectrum at the site of the NV centre shown in Fig. 4c. \textbf{c,} Calculated magnetic noise spectrum at 50 nm above an infinite magnetic plane as a function of external magnetic field $B_{\mathrm{ext}}$ (Supplementary Section VI). The increasing NV spin-state relaxation rates for decreasing $B_{\mathrm{ext}}$ observed in Fig. 4b are consistent with the calculated increase in noise spectral density at the NV ESR frequencies, which are denoted by the yellow and white dashed line for the $m_s = 0 \leftrightarrow +1$ and $m_s = 0 \leftrightarrow -1$ transition respectively.}
\label{fig:fig4}
\end{figure}

\noindent
To understand this behaviour, we calculate the magnetic-noise spectrum at 50 nm above an infinite magnetic plane (Fig. 4c, Supplementary Section VI). We use a general framework describing the noise spectrum at the site of a sensor spin in terms of the spin-spin susceptibility and a $k$-space filter function associated with the dipolar coupling to the spins in the magnet (Supplementary Section VI). This filter function contains a kernel $\sim k^2 e^{-2kd}$ which peaks at $k=1/d$, where $d$ is the NV-disc distance, illustrating that the noise at the site of the NV centre is dominated by spin-spin correlations on the scale of d and validating our infinite-plane approximation. The model excellently describes the measured increase in spin noise at the NV ESR frequencies as we lower $B_{\textrm{ext}}$ (Fig. 4b), resulting from the strong spectral content above the FMR frequency (Fig. 4c). These relaxation measurements can be extended with $T_1^{\rho}$- and $T_2$-spectroscopy techniques\cite{Rosskopf14} to characterize a spin-noise spectrum over a range of frequencies at a fixed value of magnetic field.\\

Looking ahead, the complementary NV magnetometry techniques, demonstrated here for spin waves in a ferromagnetic disc, open up exciting possibilities to explore a wide variety of magnetic excitations in nanoscopic systems under ambient conditions. We envision nanometre-scale studies of magnetic vortices and skyrmions, atomically-engineered quantum magnets, and spin ice. In addition, these techniques can be applied to characterize the magnetic fields generated by edge currents in quantum Hall systems and topological insulators\cite{Nowack13}.\\

\textbf{Methods}\\
\textit{Application of $B_{\textrm{ext}}$.} We apply the static external field $B_{\textrm{ext}}$ along the axis of target NV centres to assure good optical spin contrast (Supplementary Section II). $B_{\textrm{ext}}$ is thus oriented at an angle of 54$^{\circ}$ w.r.t the plane of the disc. Throughout this work, we select NV centres with equally oriented crystal axes. To avoid hysteresis in the disc, in all measurements we first apply a large field ($B_{\textrm{ext}} >$ 700 G) and then sweep the field down in small steps. \\
\textit{DC magnetometry.} The ESR frequencies of an NV centre in a magnetic field $\mathbf{B}$ are determined by the Hamiltonian $\mathscr{H} = D S_z^2 + \gamma \mathbf{B} \cdot \mathbf{S}$ where $S_{i=x,y,z}$ are Pauli spin matrices for a spin 1, and $z$ denotes the direction of the NV centre crystal axis. We use this Hamiltonian to calculate the projection of the magnetic field onto the NV-axis from the measured ESR frequencies (Fig.1c-d), as described in detail in Supplementary Section II.\\
\textit{Normalization procedure.} To obtain the signal in Fig.\ 3, we apply the pulse sequence in Fig.\ 3a and normalize the PL upon spin readout using two reference measurements. In these measurements, we apply the sequence of Fig. 3a without the MW drive field and with the final $\pi/2$-pulse around the x or the -x axis which yields minimum and maximum PL values. Using these bounds we normalize the PL measured at the end of the pulse sequence in Fig. 3a to obtain $B_{\textrm{eff}}$ (Supplementary Figure 3). We then divide $B_{\textrm{eff}}$ by the square of the driving field $|b_D|^2$, which we independently determine by measuring the Rabi frequency of NV$_{\mathrm{ref}}$ as a function of the ESR frequency (Supplementary Figure 4). The measured linear scaling of $B_{\textrm{eff}}$ with the MW-source power validates this normalization procedure (Supplementary Figure 5). The normalization is described in detail in Supplementary Section IV.\\

\noindent
\textbf{Acknowledgements} 
We acknowledge support of the DARPA QuASAR program and the National Science Foundation (NSF). F.C. acknowledges support from the Swiss National Science Foundation (SNSF). 

\begingroup
\renewcommand{\addcontentsline}[3]{}

\endgroup

\pagebreak
\widetext
\begin{center}
\textbf{\large Supplementary Information: Nanometre-scale probing of spin waves using single electron spins}
\end{center}
\setcounter{equation}{0}
\setcounter{figure}{0}
\setcounter{table}{0}
\setcounter{page}{1}
\makeatletter

\tableofcontents

\section{Setup and sample}
\noindent
Our experiments were performed on a type IIa diamond grown by chemical vapor deposition (by the company Element 6) measuring 2x2x0.05mm$^3$. We studied NV centres formed by N$^{15}$ ion implantation at 18keV and a density of $30/\mu$m$^2$ and subsequent annealing for 2 hours at 800$^\circ$C, yielding NV centres at an estimated $\sim$50 nm below the diamond surface.

\noindent
The magnetic fields used to control the NV centre spin state and to drive spin-wave excitations in the disc were generated by microwave (MW) currents. These currents were delivered to the sample by a Ti/Au 5nm/95nm coplanar waveguide (CPW) with a gap size of 15$\mu$m fabricated on the surface of the diamond. In the centre of the CPW gaps, we fabricated permalloy (Ni$_{81}$Fe$_{19}$) microdiscs by e-beam lithography and subsequent e-beam evaporation. The diamond was glued on a coverslip using optically transparent wax and NV centres were optically accessed through the coverslip-diamond stack with an NA=1.25 oil-immersion objective, resulting in the geometry depicted in Fig. 1a of the main text. 

\noindent
Laser pulses for optical spin initialization and readout were generated by an acousto-optical modulator (AOM) in double-pass configuration. The NV centre spin state was read out by integrating the first 550 ns of photoluminescence during a laser pulse. To avoid melting of the Permalloy disc, we limited the laser power to ~250 $\mu$W.

\noindent
The MW bursts used to control the NV spin state and to excite spin-waves in the disc were generated by modulating the output of a MW source (Rohde\&Schwarz SMB100a or Agilent N5183A) using pulses generated by an arbitrary waveform generator (Tektronix AWG520) input into a MW switch (Minicircuits ZASWA-2-50DR+ or RFLambda RFSPSTA0208G). To control the NV centre along two orthogonal axes (see Fig. 3a of main text), an additional IQ mixer was incorporated (Marki IQ1545). After amplification (Minicircuits ZVE-3W-83+ 2-8 GHz and ZHL-16W-43+ 1.8-4 GHz), signals from different MW sources were combined using a diplexer (K\&L microwave 5IZ10-3050) and fed to the on-diamond CPW through a printed circuit board.

\noindent
For all the experiments, a static magnetic field was applied that separates the two NV centre spin transitions, which allowed the individual addressing of a target transition by MW pulses at the associated ESR frequency. To assure good optical spin contrast, the static field was carefully aligned with the NV centre crystal axis, using a procedure described in the next section.

\section{Photoluminescence-based field alignment procedure}
\begin{figure}[h!]
\includegraphics[width=1\textwidth]{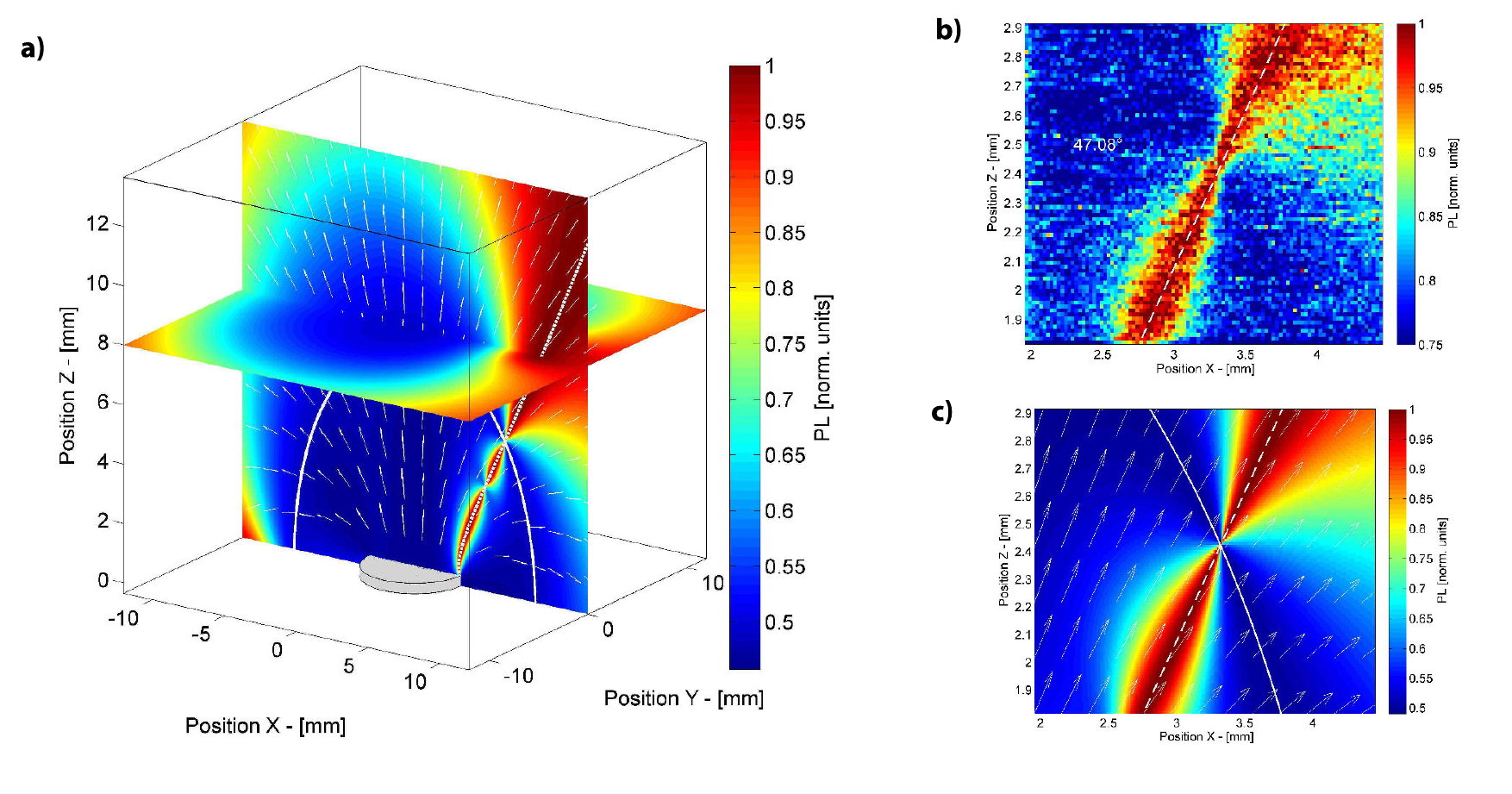}
\caption{\textbf{NV-centre photoluminescence (PL) as a function of magnet position.} Left: Numerically calculated PL for an NV centre in the field of a cylindrical NdFeB permanent magnet of diameter 6.35 mm and height 12.7 mm (the model magnet D48-N52 produced by K\&J magnetics). The NV axis is oriented along the $[\sin(\theta_{\text{NV}}), 0, \cos(\theta_{\text{NV}})]$ direction, with $\theta_{\text{NV}} = \arccos(1/\sqrt{3})$, and moved in space in the vicinity of the magnet's surface (gray disc). The origin of the reference frame represents the centre of the magnet's top surface. The white solid line marks the positions where the field is 514 G. The dashed white line marks the positions where the field is aligned with the NV axis. White arrows represent the directions of the field in the various positions. Right: Experimentally measured (top) and numerically computed (bottom) photoluminescence (PL) in the vicinity of the $\sim 500$ G region. There, the PL is easily quenched because of the spin mixing corresponding to the level anti-crossing of the NV excited states.\cite{Tetienne12} The calculation/measurement refers to a NdFeB permanent magnet of diameter 3.175 mm and height 9.525 mm (model magnet D26-N52 produced by K\&J magnetics).}
\label{fig:PL}
\end{figure}
\noindent
In our experiments, we use the spin of the NV centre as an optically interrogated magnetometer. To assure good optical spin contrast, it is essential to align the applied magnetic field $B_{ext}$ with the N-V crystal axis \cite{Tetienne12}. In our experiments, all field sweeps are conducted by translating a permanent magnet along computed space trajectories that keep $B_{ext}$ aligned with the NV axis. In this section, we describe the procedure we used to find these trajectories, based on combining a model of the field generated by our magnet with a model of the NV-centre photoluminescence described by Tetienne et al\cite{Tetienne12}. In section \ref{Test1} we describe experimental tests of the quality of the alignment.   
\subsection{Calculation of the photoluminescence}
\label{CalcI}
\noindent
To apply $B_{ext}$, we used cylindrical NdFeB magnets of variable dimensions. We calculate their space-dependent magnetic field profile with the open source Radia package\cite{Radia97}. Furthermore, we calculate the PL of an NV centre in a field using the rate-equation model and transition rates from Ref.~\onlinecite{Tetienne12}. Combining these models yields characteristic plots for the space-dependent PL (Fig.~\ref{fig:PL} left panel). The white dashed line marks the trajectory corresponding to perfect alignment. It is clear that the PL is very sensitive to misalignment whenever the field is $\approx 500$ G or $\approx 1000$ G. These field values correspond to the level anti-crossing of the ground and excited NV states, respectively.\cite{Tetienne12}. A direct comparison with experimental data around $\approx 500$ G (Fig.~\ref{fig:PL}, right panel) illustrates  the quality of the model. 

\noindent
A transformation between the laboratory frame and the model frame is essential for calculating the magnet position corresponding to a given target magnetic field. The procedure to obtain this transformation is described next.
\subsection{Determining the magnet trajectory along which the field is well aligned}
\label{WahbaI}
\noindent
To calculate the magnet trajectory along which the field is well aligned with the NV axis, we use the strong PL dependence on the magnet position in the $B_{ext} \geq 514$ G field range, which can be easily identified by the nodal point in the PL (Fig.~\ref{fig:PL} right panel). In particular, we first translated the magnet along orthogonal directions lying within $N$ different constant-$Z$ planes, recording the coordinates of a set of $\mathbf{y}_i$ points, $\{i=1,...,N\}$, where the PL was found to peak. The $\mathbf{y}_i$ represent a set of magnet positions in the laboratory frame for which the field is aligned. For each of these $\mathbf{y}_i$ points we measured $B_{ext,i}$ by ESR. Provided with $B_{ext,i}$, we numerically compute the corresponding $\mathbf{x}_i$ positions of the magnet in the model frame. The transformation relating the $\mathbf{y}_i$ to the $\mathbf{x}_i$ is:
\begin{equation}
\mathbf{y}_i = \mathbf{R} \mathbf{x}_i + \mathbf{t},
\label{eq:Waha1}
\end{equation}
where $\mathbf{R}$ is assumed to be a pure rotation matrix and $\mathbf{t}$ a translation vector. We obtain the matrix $\mathbf{R}$ via Wahba's method\cite{Landis99}, which provides an expression for $\mathbf{R}$ based on the minimization of the following cost function:
\begin{equation}
\mathscr{L}(\mathbf{R}) = \frac{1}{2(N-1)} \sum_{i=1}^{N-1} \left|\left| \left( \mathbf{y}_i - \mathbf{y}_1 \right) -  \mathbf{R} \left( \mathbf{x}_i - \mathbf{x}_1  \right) \right| \right|^2.
\label{eq:miniWha}
\end{equation}
The algorithms for minimizing $\mathscr{L}(\mathbf{R})$ and computing $\mathbf{R}$ are described in Ref.~\onlinecite{Landis99}. We then obtain the vector $\mathbf{t}$ by:
\begin{equation}
\mathbf{t} = \frac{1}{N} \sum_{i=1}^N \left( \mathbf{y}_i - \mathbf{R}\mathbf{x}_i  \right).
\label{eq:trasl}
\end{equation}
Provided with the matrix $\mathbf{R}$ and the vector $\mathbf{t}$, the laboratory frame coordinates of the magnet can be calculated for any value of $B_{ext,i}$ via \eqref{eq:Waha1}. The next section describes experimental tests of the quality of the field alignment by measuring the field dependence of the ESR resonances of NV$_{\text{ref}}$. 
\subsection{Test of the field alignment along the numerically calculated magnet trajectory}
\label{Test1}
\begin{figure}[h!]
\centering
\includegraphics[width=\textwidth]{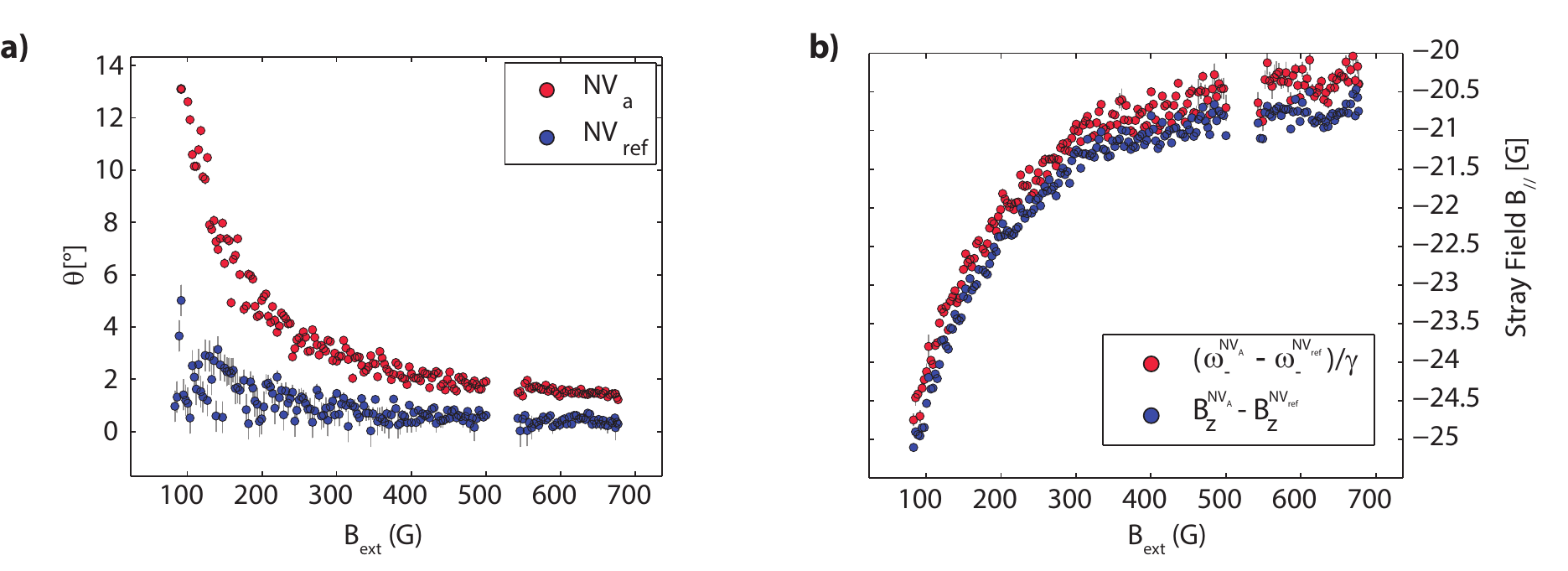}
\caption{\textbf{Test of the field alignment. a,} Misalignment angle $\theta = \arctan(B_{\perp}/B_z)$, evaluated for both NV$_{\text{ref}}$ and NV$_{\text{a}}$. \textbf{b,} Stray field $B_z^{\text{NV}_{\text{a}}}-B_z^{\text{NV}_{\text{ref}}}$ at the NV$_{\text{a}}$ (see text), as a result of the presence of the Permalloy disc. The correction for the spatial field gradient from the permanent magnet is not included in the plot.}
\label{fig:avgFr}
\end{figure}
\noindent
In order to evaluate the quality of the field alignment along the numerically calculated maganet trajectory, we use single-NV vector magnetometry\cite{Balasubramanian08}. We measure the ESR frequencies $\omega_\pm$ of the $m_s = 0 \leftrightarrow \pm 1$ transitions to determine the magnitude of the external field along ($B_z$) and transverse to ($B_\perp$) the NV axis. \\
In particular, we consider the NV Hamiltonian:
\begin{equation}
\mathscr{H} = D(\hat{S_z})^2 + \gamma B_z \hat{S_z} + \gamma B_{\perp} \hat{S_x},
\label{eq:HaminField}
\end{equation}
where $D$ is the zero-field splitting, and $\gamma = 2\pi \cdot 2.8025$ MHz/G is the NV gyromagnetic ratio.\\
The eigenvalues of eq. \eqref{eq:HaminField} are given by the solution of the characteristic equation:
\begin{align}
& || \mathscr{H} - \lambda_i \mathbf{I}  || = 0, \label{eq:char1} \\
 \lambda_1 = \lambda_0 \qquad & \lambda_2 = \lambda_0 + \omega_+ \qquad \lambda_3 = \lambda_0 + \omega_-.  \label{eq:char2}
\end{align}
We substitute eq. \eqref{eq:char2} into eq. \eqref{eq:char1} and solve for $\lambda_0,B_z,B_{\perp}$, using the measured values for $\omega_{\pm}$. We obtain: 
\begin{align}
&B_z = \frac{\sqrt{-(D + \omega_+ - 2\omega_-)(D+\omega_--2\omega_+)(D+\omega_-+\omega_+)}}{3\gamma\sqrt{3D}}, \label{eq:hz} \\
&B_{\perp} = \frac{\sqrt{-(2D-\omega_+-\omega_-)(2D+2\omega_--\omega_+)(2D-\omega_-+2\omega_+)}}{3\gamma\sqrt{3D}}. \label{eq:hx}
\end{align}
The previous expressions rely on the value of $D$, which we extracted from the field-independent average $D = (\omega_++\omega_-)/2$ to be $D = 2\pi\cdot 2.8707(1)$ GHz. In Fig. \ref{fig:avgFr}a we plot the extracted misalignment angle, defined as $\arctan(B_{\perp}/B_z)$, for both NV$_{\text{ref}}$ and NV$_{\text{a}}$. A $\theta \neq 0$ value for NV$_{\text{ref}}$ is solely due to a misaligned field of the permanent magnet, which is limited to the small value of $\theta \leq 3^{\circ}$  (Fig. \ref{fig:avgFr}a). For NV$_{\text{a}}$, the stray field from the Py disc causes the low-field increase of $\theta$. 

\subsection{Determining the disc field from the ESR traces}
\noindent
The field created by the permanent magnet varies slightly between the different NVs studied in this work. In addition, the associated field gradient varies also with the magnitude of the external static field applied along the NV centre axis. By measuring the externally applied field at reference NV sites distant from the Permalloy disc, we measure a field gradient of $\sim 0.2$ G/$\mu$m at the maximum static field used in experiments of $B_z^{\text{NV}_{\text{ref}}} = 700$ G (applied using the model magnet D48-N52 produced by K\&J magnetics). We confirm numerically the strength and field-dependence of the measured field gradient, and the fact that the latter is uniform in space within the optical field of view. In this way, we can compute the field-dependence $|\Delta B_z^{\text{NV}_{\text{a,b}}} |$ of the  difference between the external static field at the NV sites $a,b$ with respect to the reference NV.\\
\noindent
Finally, using the methods described in the previous paragraph, the values of $B_z$ at both NV$_{\text{ref}}$ and NV$_{\text{a,b}}$ as extracted from the ESR resonances allow us to evaluate the residual stray field $B_{//} = B_z^{\text{NV}_{\text{a,b}}}-B_z^{\text{NV}_{\text{ref}}} + |\Delta B_z^{\text{NV}_{\text{a,b}}} |$ created by the Permalloy disc only. In \fref{avgFr}b we compare $B_z^{\text{NV}_{\text{a,b}}}-B_z^{\text{NV}_{\text{ref}}}$ with the simpler stray field estimate $(\omega_-^{\text{NV}_{\text{ref}}}-\omega_-^{\text{NV}_{\text{a}}})/\gamma$. In the main text, we show the field dependence of $B_{//}$, which includes the correction for the field-gradient.

\section{Model for the Fano interference in the Rabi oscillations}
\noindent
In this section we describe the fitting procedure for the Fano resonances observed in Fig.\ 2 of the main text. We use a phenomenological model by considering the interference between the response of a linear system and its driving force. In particular, we use the analogy between the response of a ferromagnet to a driving field and the response of a simple mechanical harmonic oscillator. In both cases energy is stored at resonance; the coherent precession of the two spin components transverse to the equilibrium axis of the ferromagnet is analogous to the periodic transformation of kinetic into potential energy. Furthermore, energy dissipation occurs via spin damping in the ferromagnet and through friction in the oscillator. Finally, both systems are characterized by a phase delay in their response upon driving. Let's consider the equation of motion for a simple mechanical oscillator:
\begin{equation}
m \ddot{x}(t) + \beta \dot{x}(t) + m \Delta^2 x(t) = F(t),
\label{eq:harm}
\end{equation}
where $m$ is the particle's mass, $F(t)$ the driving force, $\beta$ the friction, $\Delta$ the resonance frequency and $x(t)$ the time-dependent coordinate. The oscillator's response in frequency space reads as:
\begin{equation}
x(\omega) = \chi(\omega) F(\omega)= \frac{F(\omega)}{m (\Delta^2 - \omega^2) + i \beta \omega},
\label{eq:harm2}
\end{equation}
where $\chi(\omega)$ is the dynamical susceptibility. Similarly, the transverse dynamical magnetic uniform susceptibility $\chi_{\perp}(\omega)$ for a ferromagnet with short-range exchange interactions reads:\cite{Trifunovic13}
\begin{equation}
\chi_{\perp}(\omega) = \frac{g(S)}{(\Delta-\omega) + i W},
\label{eq:harm3}
\end{equation}
where $W$ is the FMR line width, $\Delta$ the FMR frequency and $g(S)$ a constant prefactor which depends on the spin quantum number. A Fano interference is created in the mechanical system described by \eqref{eq:harm2} when considering the total response:
\begin{equation}
x_{\mathrm{tot}}(\omega) = \left( 1 + \chi(\omega) \right) F(\omega) = \left( 1 + r(\omega) e^{i \theta(\omega)} \right) F(\omega),
\label{eq:Fano}
\end{equation}
which implies a quadrature summation of the total normalized output's amplitude:
\begin{equation}
\left| \frac{x_{\mathrm{tot}}(\omega)}{F(\omega)} \right| = \sqrt{ 1 + r^2(\omega) + 2 r(\omega) \cos(\theta(\omega))}.
\label{eq:Fanotot}
\end{equation}
Eq. \eqref{eq:Fanotot} is the mechanical equivalent of the expression we have used to model the enhanced normalized Rabi frequency. In particular, in the magnetic case, by calling the total transverse field at the NV $i$-site $b^i(\omega)$ and the external driving field $b_D(\omega)$ we obtain:
\begin{equation}
\left| \frac{b^i(\omega)}{b_D(\omega)} \right| = \sqrt{ 1 + r^2(\omega) + 2 r(\omega) \cos(\theta(\omega) + \phi_i)},
\label{eq:FanototMagn}
\end{equation}
where $\phi_i$ is an additional frequency-independent phase factor motivated by the fact that the stray field created by the ferromagnet varies according to the different NV $i$-site. The terms $\theta(\omega)$ and $r(\omega)$ can be obtained by taking argument and norm of the complex susceptibility in eq. \eqref{eq:harm3}. It is important to note that in order to model our field-dependent normalized Rabi curves (main-text Fig. 2b), one has to impose for the FMR resonance the Kittel-like expression $\Delta = \gamma \sqrt{B_{\mathrm{ext}} (B_{\mathrm{ext}} + A)}$ and for the frequency $\omega = D + \gamma \left( B_{\mathrm{ext}} + B_{\parallel}(B_{\mathrm{ext}}) \right)$, where $D$ is the zero field splitting and $B_{\parallel}$ is the projection along the NV axis of the ferromagnet's static stray field (as measured in main-text Fig. 1c). 
\section{Normalization procedure for the off-resonant detection scheme (Fig.\ 3 of main text)}
\noindent
In this section we describe the normalization procedure used to obtain Figs.\ 3b,d of the main text. 

\noindent
We use the measurement scheme shown in Fig. 3a of the main text, in which we apply the final $\pi/2$-pulse along the x- or y-axis and subsequently read out the spin-dependent PL ($P_x$ and $P_y$ resp.). During the same measurement, we also apply two normalization sequences which are the same as in Fig. 3a of the main text except that we turn the MW off and apply the final $\pi/2$-pulse along the $x$ and $-x$ axis. \fref{Normalization}a shows the raw data of these four measurements. The normalization sequences yield the minimum and maximum PL ($P_{min}$ and $P_{max}$ resp.) which we use to obtain the x and y spin expectation values (\fref{Normalization}b) according to $\langle i \rangle= 2\frac{P_i - P_{min}}{P_{ax} - P_{min}}-1$, where $i=x,y$.

\noindent
From the expectation values, we calculate the phase $\phi$ of the superposition using $\phi = \arctan(\frac{\langle y \rangle}{\langle x \rangle})$ (\fref{Normalization}c). We express the final signal in terms of an effective field $B_{eff}=\phi/(\gamma T)$, and divide by the square of the driving field $|b_d|^2$ to correct for the frequency-dependent delivery of microwaves through our setup (\fref{Normalization}d). We independently characterized $|b_d|^2$ by measuring the Rabi frequency of NV$_{ref}$ at constant MW-source power as a function of the ESR frequency, which we tune using $B_{ext}$ (\fref{RabiCal}). The inset of \fref{Normalization}c illustrates the effect of the frequency-dependent power on the measurement of $\phi$: dips/peaks in $\phi$ clearly correspond to dips/peaks in the Rabi frequency, motivating the normalization by $|b_d|^2$.

\begin{figure}[h!]
\begin{center}
\includegraphics[width=\textwidth]{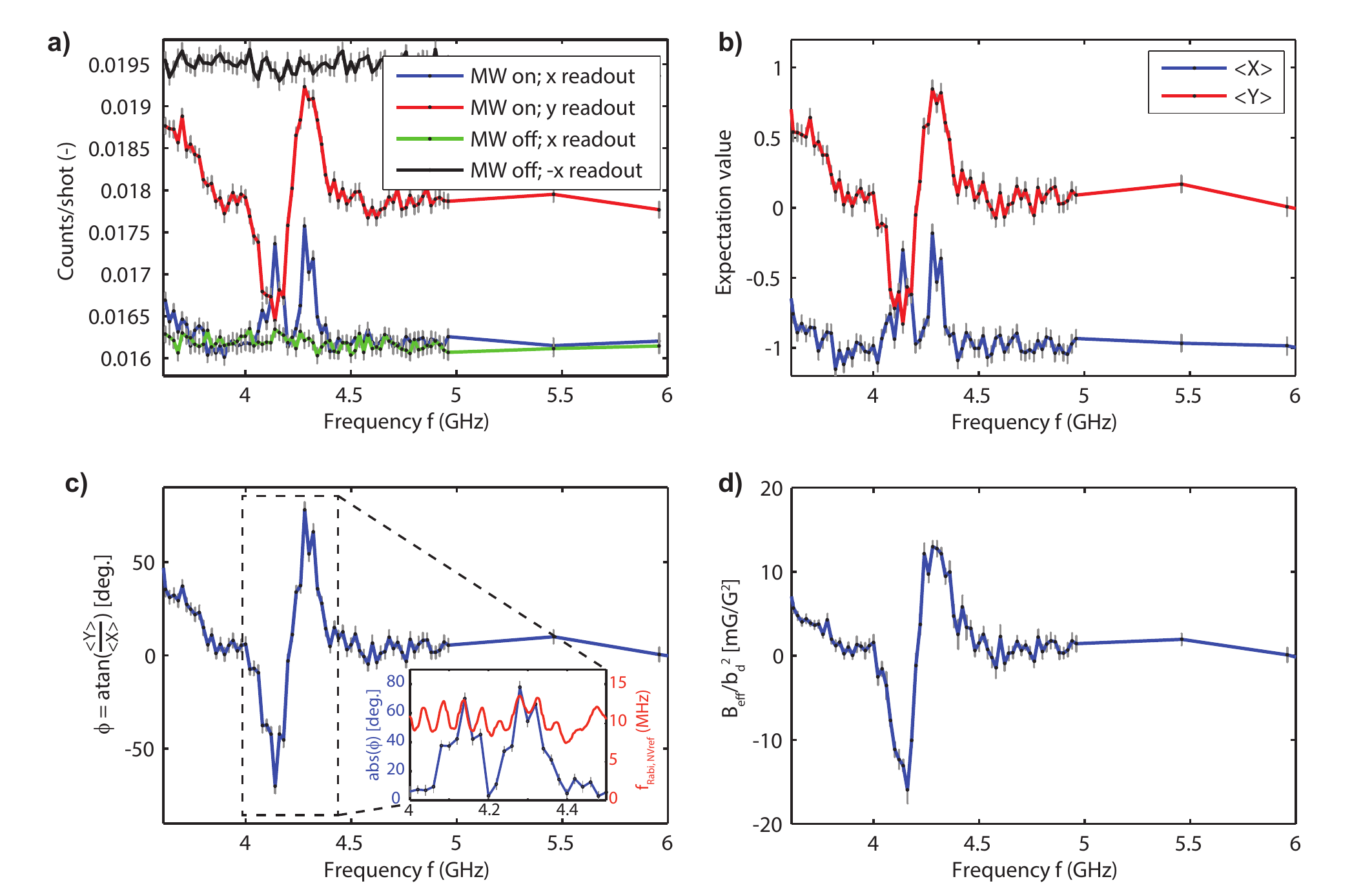}
\end{center}
\caption{\label{fig:Normalization} \textbf{Example of the normalization procedure used to obtain main-text Fig. 3b and 3d: a frequency sweep at $B_{ext}=211$ G on NV$_A$. a,} Raw data (details on pulse sequences in text) \textbf{b,} Spin expectation values, \textbf{c,} Phase of the final superposition. Inset: Absolute value of the phase and Rabi frequency of NV$_{ref}$ plotted in one graph. The dips/peaks in Rabi frequency clearly correspond to dips/peaks in $\phi$. \textbf{d,} Effective field normalized by the power of the driving field.}
\end{figure}
\begin{figure}[h!]
\begin{center}
\includegraphics[width=\textwidth]{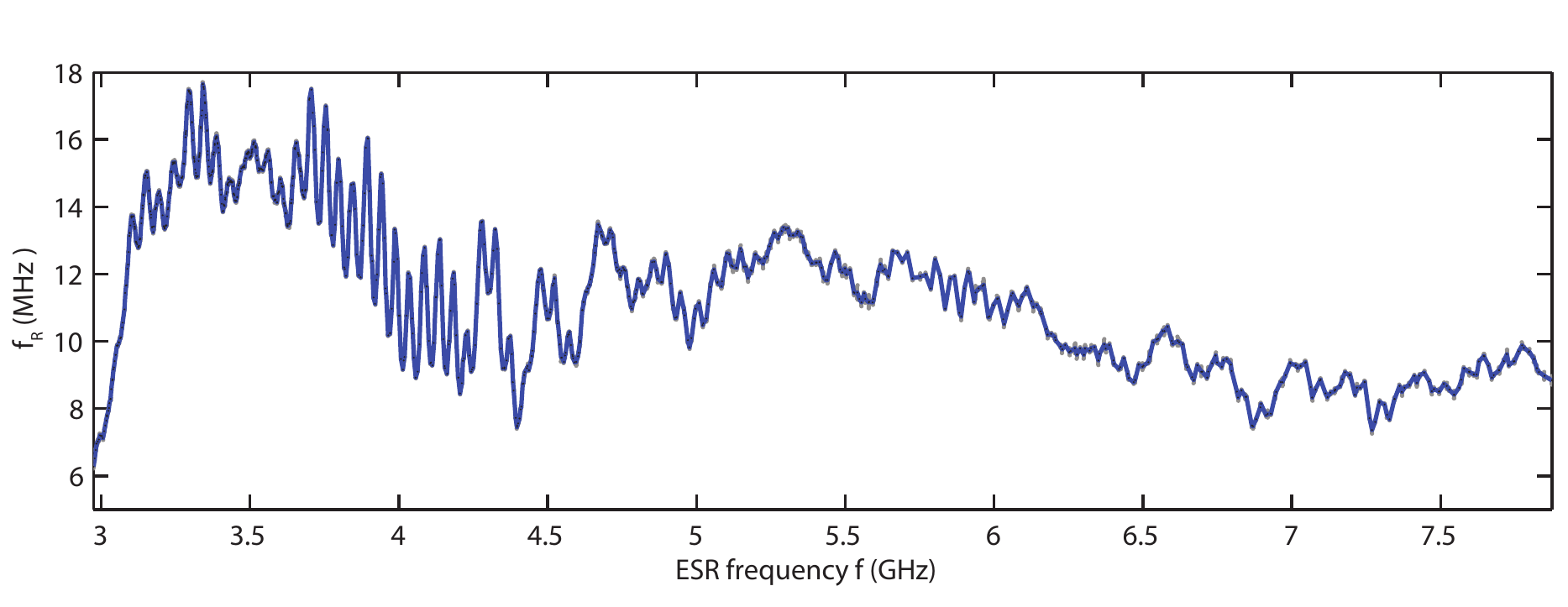}
\end{center}
\caption{ \textbf{Characterization of the frequency-dependent transmission of microwaves through our setup.} Even though this measurement is performed with constant MW-source power, we observe a strongly frequency-dependent spin rotation rate (Rabi frequency $f_R$). Measuring the Rabi frequency is an excellent method to characterize in-situ how much MW power actually reaches the NV centre. From the Rabi frequency $f_R$, we obtain $|b_d| = f_R/\gamma$. Measurement performed on NV$_{ref}$, the ESR frequency is tuned using $B_{ext}$.}
\label{fig:RabiCal}
\end{figure}

\noindent
To validate the procedure of normalizing $B_{eff}$ by the square of the drive field $|b_d|^2$ (\fref{Normalization}d), we studied the dependence of $B_{eff}$ on the power of the MW source $R$, since $|b_d|^2 \propto R$. \fref{FMR_vsSourcePower_Full} shows the linear scaling of the signal with MW power.

\begin{figure}[h!]
\begin{center}
\includegraphics[width=\textwidth]{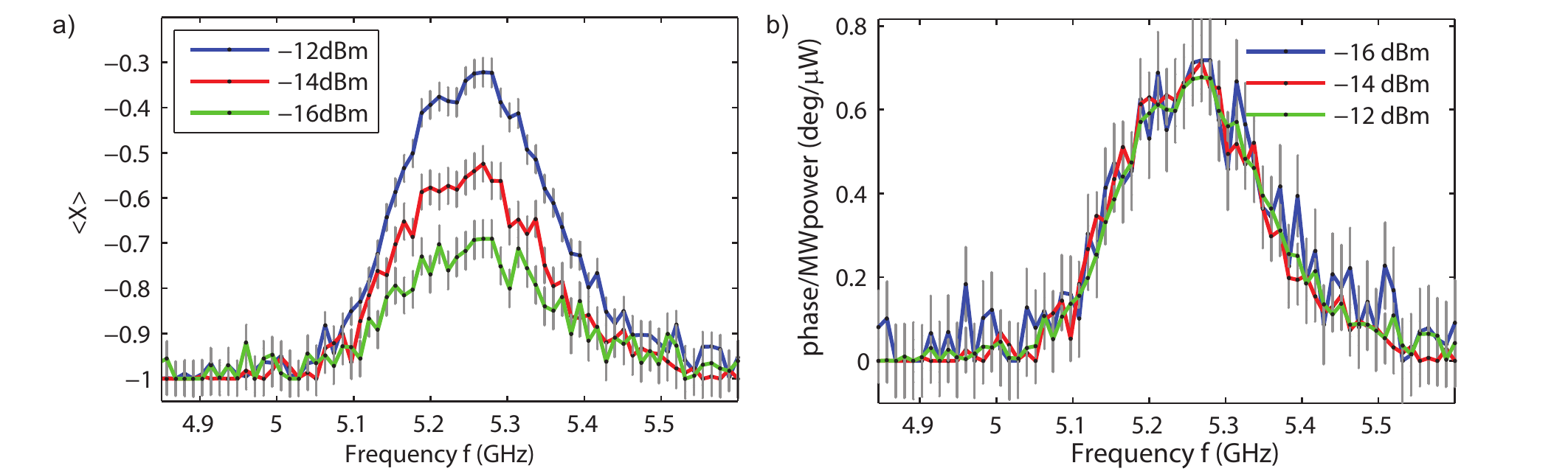}
\end{center}
\caption{ \textbf{Linear scaling of phase $\phi$ with MW-source power. a,} FMR measured on NV$_B$ for different MW powers at $B_{ext}=450$ G. \textbf{c,} The  phase divided by the MW-source power, illustrating the linear scaling with MW-source power. }
\label{fig:FMR_vsSourcePower_Full}
\end{figure}
\section{AC Stark effect in off-resonant detection scheme (Fig.\ 3 of main text)}
\noindent
In this section, we analyze the phase shift imparted on the NV spin state by the AC Stark effect caused by the off-resonant driving field in the measurement scheme of Fig. 3A of the main text. The goal is to estimate if the Stark shift contributes significantly to the measured $B_{eff}$ in Fig. 3 of the main text, taking into account that the AC drive field can be enhanced by the AC field generated by the ferromagnet as was shown in main-text Fig.\ 2.

\noindent
The Stark shift, and the related Bloch Siegert shift, \cite{Cappellaro09,London14} is usually treated within an effective two-level model.\cite{Rohr14,Cappellaro09} However, in our experiments, we prepare the $S=1$ spin of the NV centre in a superposition of $m_s=0,-1$ states, while the off-resonant driving has a frequency in the vicinity of the $m_s=0 \leftrightarrow +1$ transition. Therefore, the three-level nature of the system has to be retained.\\
To discuss the effective field which is picked up by the NV centre as a result of the Stark shift, we employ the time-dependent Schrieffer-Wolff formalism.\cite{Wang14}
The method outlined below can be readily generalized to other multilevel quantum systems. 
\subsection{Time-Dependent Schrieffer-Wolff formalism}
\label{SW:form}
\noindent
The time-dependent Schrieffer-Wolff formalism is discussed in detail in Ref.~\onlinecite{Wang14}. Here we briefly recall the main results.
We assume an Hamiltonian composed of two parts, a time-independent diagonal part $H_0$ and a time-dependent non-diagonal part $\mathscr{H}_{nd}(t)$:
\begin{equation}
\mathscr{H}(t) = \mathscr{H}_0 + \mathscr{H}_{nd}(t). \label{eq:ham0}
\end{equation}
In order to derive an effective Hamiltonian $\mathscr{H}_{eff}$ which retains up to second order the effects of the perturbation $\mathscr{H}_{nd}(t)$, we change the quantum basis by applying a unitary transformation $\hat{U}(t) = e^{\hat{S}(t)}$, with $\hat{S}(t) = -\hat{S}(t)^{\dagger}$.\\
In the Schrieffer-Wolff formalism, a series expansion of $\hat{S(t)}$ leads to the expression for $\mathscr{H}_{eff}$, which reads:\cite{Wang14}
\begin{equation}
\mathscr{H}_{eff} = \mathscr{H}_0 + \frac{1}{2} \left[ \hat{S}(t), \mathscr{H}_{nd} \right] + O(\mathscr{H}_{nd}^3).
\label{eq:HeffSW}
\end{equation}
The matrix $\mathbf{S}(t)$ can be computed from the condition that eliminates the first-order terms $O(\mathscr{H}_{nd})$, namely:\cite{Wang14}
\begin{equation}
\mathscr{H}_{nd}(t) + \left[ \hat{S}(t), \mathscr{H}_0 \right] + i \hat{\dot{S}}(t) = 0.
\label{eq:swcond}
\end{equation}
We will now use this formalism to obtain an expression for the time-independent part of $\mathscr{H}_{eff}$ for an NV centre in an off-resonant driving field.
\subsection{Description of the AC Stark shift for an NV centre in an off-resonant driving field}
\label{SW:NVdriv}
\noindent
For an NV centre in a longitudinal static field $B_z$ and a transverse dynamic field $B_2$, the Hamiltonian reads:
\begin{equation}
\mathscr{H}(t) = \mathscr{H}_0 + \mathscr{H}_{nd}(t) = D(\hat{S}^z)^2 + h \hat{S}^z + h_2 \cos(\omega t) \hat{S}^x, \label{eq:hamNV}
\end{equation}
where $S=1$ spin matrices are used, and we have defined $h=\gamma B_z$ and $h_2 = \gamma B_2$. We assume for $\hat{S}(t)$ the following 3x3 matrix representation, which satisfies the $\hat{S}(t) = -\hat{S}(t)^{\dagger}$ condition:
\begin{align}
\hat{S}(t) =  \left( \begin{array}{ccc} 0	& {S}_1(t) & 0 \\ -{S}^*_1(t) & 0 & {S}_2(t) \\ 0 & -{S}^*_2(t) & 0 \end{array} \right). 
\label{eq:Srepre}
\end{align}
Provided with eq. \eqref{eq:swcond}, the previous expression for $\hat{S}(t)$ defines a set of two linear non-homogeneous  differential equations: 
\begin{align}
& i \dot{S}_1(t) - \omega_+ S_1(t) + h_2 \cos(\omega t) \frac{\sqrt{2}}{2} = 0 
& i \dot{S}_2(t) + \omega_- S_2(t) + h_2 \cos(\omega t) \frac{\sqrt{2}}{2} = 0  
\label{eq:eqs}
\end{align}
where we have defined $\omega_{\pm}=D\pm h$, which are the spin transition frequencies in the absence of the Stark shift. We compute the solution to eq. \eqref{eq:eqs} imposing $S_{1,2}(t=0)=0$, i. e.,  $\hat{U}(t=0) = \hat{I}$. We obtain:
\begin{align}
S_1(t) =  - h_2 \frac{ \omega_+ e^{-i\omega_+ t} - \omega_+ \cos(\omega t)  + i \omega \sin(\omega t)  }{\sqrt{2} (\omega_+ - \omega)(\omega_+ + \omega)} \nonumber \\
S_2(t) = h_2 \frac{ \omega_-  e^{-i\omega_- t }  - \omega_- \cos(\omega t)  - i \omega \sin(\omega t)}{\sqrt{2} (\omega_- - \omega)(\omega_-+\omega)}
\label{eq:s1s2}
\end{align}
By using eq. \eqref{eq:HeffSW}, the effective Hamiltonian can be written as:
\begin{equation}
\mathscr{H}_{eff} \approx \mathscr{H}_0 + \frac{1}{2} \left[ \hat{S}(t), \mathscr{H}_{nd} \right] = 
\mathscr{H}_0 +  h_2 \cos(\omega t) \frac{\sqrt{2}}{2} \left(  \begin{array}{ccc} \frac{S_1(t) + {S}^{\dagger}_1(t)}{2}	& 0 & \frac{S_1(t)-S_2(t)}{2} \\ 0 & \frac{S_2(t) + {S}^{\dagger}_2(t)}{2} - \frac{S_1(t) + {S}^{\dagger}_1(t)}{2} & 0 \\ \frac{{S}^{\dagger}_1(t) - {S}^{\dagger}_2(t)}{2} & 0 & -\frac{S_2(t) + {S}^{\dagger}_2(t)}{2} \end{array} \right).
\label{eq:effs1d2V2}
\end{equation}
The resulting time-independent part $\mathscr{\bar{H}}_{eff}$ of the Hamiltonian \eqref{eq:effs1d2V2} has the following form:
\begin{align}
\mathscr{\bar{H}}_{eff} &= \mathscr{H}_0 + \left(  \begin{array}{ccc} A & 0 & \frac{A-B}{2} \\ 0 & B-A & 0 \\ \frac{A-B}{2} & 0 & -B \end{array} \right), \qquad
A = \frac{h_2^2 \omega_+}{4(\omega_+-\omega)(\omega_++\omega)} \qquad B = \frac{-h_2^2 \omega_-}{4(\omega_--\omega)(\omega_-+\omega)}.
\label{eq:ABNVform}
\end{align}
Note that this effective Hamiltonian contains, in second order, off-diagonal terms that couple directly the $m_s=\pm 1$ subspace. However, in our experiments, the degeneracy of the $m_s = \pm 1$ states is lifted by the static external field $h$, which represents the dominant energy scale of the subspace. For this reason, as long as $h \gg h_2$, the spin dynamics will mainly be governed by the diagonal elements of $\mathscr{\bar{H}}_{eff}$. In the limit $h \gg h_2$ we are therefore allowed to write:
\be
\mathscr{\bar{H}}_{eff}  \approx \mathscr{H}_0 + \left(  \begin{array}{ccc} A & 0 & 0\\ 0 & B-A & 0 \\ 0 & 0 & -B \end{array} \right) \nonumber \\
\label{eq:ABNV2}
\ee
which describes the shift of the energy levels of the NV spin in the presence of an off-resonant driving field.
\subsection{Analysis of the Stark effect in the off-resonant detection scheme}
\noindent
We now consider the magnetometry sequence shown in Fig. 3a of the main text. The first $\pi/2$ pulse creates a superposition of $m_s=0$ and $m_s=-1$ states, and all further NV-pulses are also given on the $m_s=0\leftrightarrow -1$ transition. The normalized contrast is given by:
\begin{align}
& C(\tau,\omega) = |\langle 0|\Psi\rangle|^2 \nonumber \\
& = \langle 0 |U_{y}(\pi/2) \exp(-i \mathscr{H}_0 \tau/2) U_{y}(\pi) \exp(-i \mathscr{\bar{H}}_{eff}(\omega) \tau) U_{y}(\pi) \exp(-i \mathscr{H}_0 \tau/2)U_{x}(\pi/2)| 0 \rangle|^2. 
\label{eq:fulltimeevo}
\end{align}
where $U_{x,y}(\pi/2)$ denotes the operator for a $\pi/2$-pulse around the $x,y$ axis.
We obtain the following result:
\begin{align}
C(\tau,\omega) = \frac{1}{2} \left( 1 + \sin \left[ 
\frac{h_2^2 \tau}{8} \left( \frac{1}{\omega_+-\omega} + \frac{2}{\omega_- +\omega} + \frac{2}{\omega_--\omega} + \frac{1}{\omega_++\omega} \right)  \right] \right)
\label{eq:fulltimeevo2}
\end{align}
For illustrative purposes, we now consider two limiting cases of small detuning. If we apply off-resonant driving near the $m_s=0\leftrightarrow +1$ transition at $\omega = \omega_+ + \delta$, similar to Fig. 3b-c of the main text, we get 
\begin{equation}
C(\tau,\omega) \underset{\omega \rightarrow \omega_+ + \delta}{\approx} \frac{1}{2} + \frac{1}{2}\sin\left(\frac{h_2^2}{8\delta} \tau \right).
\label{eq:case2C}
\end{equation}
for small $\delta$.\\
On the other hand, if we would have applied driving near the $m_s=0\leftrightarrow -1$ transition at $\omega = \omega_- + \delta$, we get
\begin{equation}
C(\tau,\omega) \underset{\omega = \omega_- + \delta}{\approx} \frac{1}{2} +  \frac{1}{2} \sin\left(\frac{h_2^2}{4\delta} \tau \right) .
\label{eq:case1C}
\end{equation}
which is the same as obtained from a two-level treatment of the NV-centre\cite{Cappellaro09}. In this case, the Stark shift is twice that of \eref{case2C} because the driving is applied near the transition frequency of the two states forming the superposition.
%
%
\begin{figure}[th!]
\centering
\includegraphics[width=\textwidth]{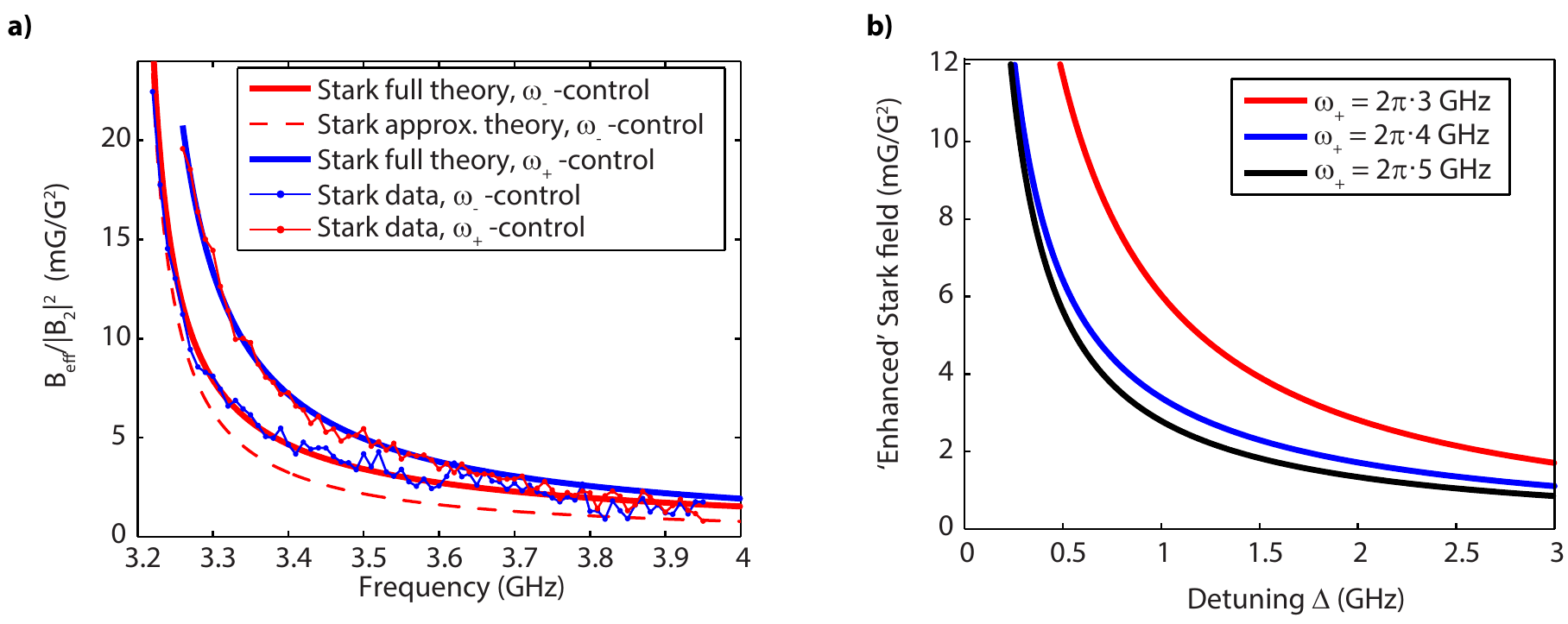}
\caption{\textbf{Analysis of the Stark shift, theory and experiment. a,} Effective Stark field $B_{eff}$ as a function of the frequency of the off-resonant driving field at $B_{ext} = 110$ G, where $\omega_+ = 2\pi\cdot3.19$ GHz. We used the measurement scheme depicted in Fig. 3a of the main text. The red (blue) dots correspond to NV-control pulses applied at the $0\leftrightarrow-1$ ($0\leftrightarrow+1$) transition. Solid lines represent the parameter-free theoretical prediction given by eq. \eqref{eq:fulltimeevo2}. The dashed line is the result expected from eq. \eqref{eq:case2C}. Because at this low field we have $(\omega_- - \omega) \sim (\omega_+ -\omega)$, eq. \eqref{eq:case2C} (dashed red line) does not correctly approximate the experimental results.
We normalized the effective field by the square of the drive field $B_2^2$ (see main text). 
\textbf{b,} Effective Stark field, \textit{multiplied by a factor 4}, as a function of the detuning $\Delta=\omega-\omega_+$ for NV-control pulses applied at the $0\leftrightarrow-1$ transition. The factor four corresponds to the maximum observed enhanced Rabi amplitude in Fig.2 of the main text, i.e., $\max \left[ \frac{f_R^{NV_i}}{f_R^{NV_{ref}}} (f) \right] $ . To estimate the contribution of the Stark effect at the FMR frequency, we assume that the FMR enhances the drive field by approximately the same factor for the measurements in Fig. 3b-c of the main text. As is visible in the figure, the expected Stark effect varies with splitting of the $|\pm 1\rangle$ NV state, i.e. with the value of the static applied field.}
\label{fig:Stark}
\end{figure}
\subsection{Comparing the full Stark-effect model with experiments}
\noindent
As a check of our model, we now quantitatively compare eq. \eqref{eq:fulltimeevo2} with experiments, applying the measurement scheme in Fig.3a of the main text to NV$_{ref}$. In the first experiment, we apply all NV control pulses at the $0\leftrightarrow +1$ transition. In the second experiment we apply all NV pulses on the $0\leftrightarrow -1$ transition. In both experiments, we apply
off-resonant driving close to the $0\leftrightarrow +1$ transition at a frequency $\omega = \omega_+ + \delta$, with $\delta>0$. 

\fref{Stark} shows the accumulated Stark phase $\phi$ in terms of an effective field $B_{eff} = \phi/(\gamma T)$, where $T$ is the duration of the off-resonant driving. We normalized $B_{eff}$ by the square of the amplitude of the driving field $B_2^2$ to correct for a frequency-dependence in the setup transmission. We independently measured $B_2(\omega)$ by measuring the Rabi frequency $\omega_{\mathrm{R}}(\omega)$ of NV$_{ref}$ as a function of ESR frequency $\omega$ which we vary using $B_{ext}$. Note that for an NV centre it can be shown\cite{Hanson08} that $\omega_{\mathrm{R}} = B_2/\sqrt{2}$. We find a good agreement with \eref{fulltimeevo2} (solid lines) for both experiments. We indeed observe a difference in the Stark effect according to which set of spin states we apply the NV-control pulses. The dotted line shows the approximation given in \eref{case2C}, which should be compared to the red solid line.

\subsection{Estimate of the 'enhanced' Stark effect in the measurements of Fig. 3}
\noindent
In this section, we estimate the magnitude of the Stark effect at the frequency of the FMR in Fig. 3 of the main text. The Stark effect quickly diminishes with increasing detuning, which is visible in Fig. 3c of the main text in the frequency range just above the $0\leftrightarrow +1$ transition over the entire magnetic field range. However, as we observed that the drive field may be enhanced by the spin-wave field (Fig. 2 of the main text), an 'enhanced' Stark effect may contribute to the observed resonances in Fig. 3 of the main text, provided the enhancement is strong enough. 

\noindent
For the plot in the right part of Fig.~\ref{fig:Stark}, we use the experimentally observed $\approx$doubling of the Rabi frequency and plot a Stark field four times as big as the one measured in the left part of the figure. We conclude that the Stark effect is negligible for the resonance observed on NV$_B$ (top panel of Fig. 3b of main text) which is on the order of 100 mG/G$^2$. On the other hand, for NV$_A$ (Fig. 3b of main text, bottom panel), the estimated Stark effect at the FMR is on the same scale as the signal in the low-field range $B_{ext} \sim< 300$ G. However, the negative sign of $B_{eff}$ in this range cannot result from the Stark effect (the sign will be described in section \ref{sec:StarkSign}). For $B_{ext} > 300$ G, the estimated Stark effect is much smaller than the observed signal. 

\subsection{Sign of the effective field associated with the Stark shift}
\label{sec:StarkSign}
\noindent
Finally, we discuss how the Stark shift provides a convenient way to assign a sign to the effective field measured with the scheme of Fig. 3a of the main text. 

\noindent
To see if the Stark shift corresponds to a positive or negative effective field $B_{eff}$, we need to consider the Hamiltonian in \eref{ABNV2}. As an example, we assume as in our experiments that $\omega = \omega_+ + \delta$, with $\delta>0$, which leads to  $A < 0$ while $B \approx 0$. Suppose the NV is prepared in a arbitrary superposition $\Psi(t=0) = a |0\rangle+b |-1\rangle+c |+1\rangle$. The time evolution in a field $h$ and with $A<0$ will read as:
\begin{equation}
\Psi(t) = e^{i A t} \left( a |0\rangle + b e^{- i (D - h - |A|) t} |-1\rangle + c e^{-i (D+h-2|A|) t} |+1\rangle \right). \label{eq_timeevo}
\end{equation}
It is evident from eq. \eqref{eq_timeevo} that the Stark shift corresponds to a \textit{positive} effective field if we create a superposition of the 0,-1 states (with $a,b \neq 0, c=0$) because the field $h$ has changed to $h +|A|$. This positive Stark field corresponds to the experimental case. On the other hand, the effective field is \textit{negative} for a superposition $a,c \neq 0, b=0$, because the field $h$ has changed to $h - 2|A|$. Therefore, the observation of the Stark shift, which as mentioned before is visible in Fig. 3c of the main text in the frequency range just above the $0\leftrightarrow +1$ transition, allows us determine the sign of $B_{eff}$ in both the measurements in Fig. 3b-c of the main text (these measurements used the exact same pulse sequence). 

\clearpage
\section{Stray-field characterization of magnetization and spin noise}
In this section we describe the properties of stray-field magnetometry of magnetization and spin noise which are relevant for the experiments presented in this work. In particular, we will show that the NV centre probes the spatial \textit{variations} in the magnetization on the scale of the NV-disc distance, and we will derive the model used for the calculations of the field-dependent magnetic noise spectrum at the NV-site shown in Fig. 4 of the main text. 

\subsection{Stray field of a space-dependent magnetization profile}
The opposite behaviour of the disc stray field at NV$_A$ and NV$_B$ observed in Fig. 1c of the main text is a direct result of the close proximity of the NV centres to the disc and the associated sensitivity to local variations in the magnetization. In this section we explain this in more detail.

In particular, we will describe that an NV spin at a distance $d$ from the surface of the disc is mostly sensitive to local variations in the magnetization on the scale of $d$. Intuitively, this can be easily understood: on the one hand, a homogeneously magnetized infinite plane generates no stray field. On the other hand, the stray field generated by variations in the magnetization on a scale much smaller than $d$ averages out at a distance $d$. More formally, it can be shown that the stray field $\mathbf{B}(\mathbf{r}_{0}) $ at a position $\mathbf{r}_{0} = (\boldsymbol{\rho}_0,d)$ (see \fref{FourMagn}a) produced by a certain two-dimensional spin texture $\mathbf{S}(\boldsymbol{\rho})$ is given by
\be\mathbf{B}(\mathbf{r}_{0}) = 
\frac{\Gamma}{(2 \pi)^2} \int_{\mathbf{k}} \mathscr{D}(-\mathbf{k},d) \cdot \mathbf{S}(\mathbf{k}) k e^{i k {\rho}_0 \cos(\phi_0-\phi_k)} \diff k \diff \phi_k.
\label{eq:STFCont}
\ee
Here, $\mathbf{S}(\boldsymbol{k})$ is the spatial Fourier transform of $\mathbf{S}(\boldsymbol{\rho})$, $\boldsymbol{k}$ is a 2-dimensional vector in reciprocal space, $\Gamma$ is the number of dipole moments per unit surface, and $\phi_k$ the angle between the $\mathbf{k}$-vector and $z'$ (see \fref{FourMagn}a). We see that stray-field detection works as a spatial Fourier filter, with a kernel given by: 
\begin{align}
\mathscr{D}(\mathbf{k},d) = \frac{\mu_0 \mu_{eff}}{2} e^{-d k} k  \left( \begin{array}{ccc}
1 &  i \sin(\phi_k)  &  i \cos(\phi_k)  \\
i \sin(\phi_k)  & - \sin^2(\phi_k)  & - \frac{\sin(2\phi_k)}{2} \\
i \cos(\phi_k) &  - \frac{\sin(2\phi_k)}{2}  &  - \cos^2(\phi_k) \end{array} \right). 
\label{eq:dipoDefKernel}
\end{align}

The NV-centre stray-field sensor is point-like, contrary to e.g. nano-SQUIDs or Magnetic-Resonance Force Microscopy probes. Therefore, the stray field computed with eq. \eqref{eq:STFCont}, after a proper projection along the NV axis, directly describes how spatial modulations of the local magnetization couple to the NV spin. We obtain the following general set of conclusions, in principle valid for any $\mathbf{S}(\boldsymbol{\rho})$ distribution:
\begin{itemize} 
\item All the elements in the kernel contain the term $k \exp(-d k)$, which peaks at $k=1/d$. This term defines the detection annulus of the stray-field sensor. We see that stray-field magnetometry is not sensitive to any uniform ($k=0$) magnetization and also not to magnetization variations on a scale much shorter than $d$. 
\item The in-plane stray field component orthogonal to the wavevector $\mathbf{k}$ is always zero.
\end{itemize}

\begin{figure}[th!]
\centering
\includegraphics[width=\textwidth]{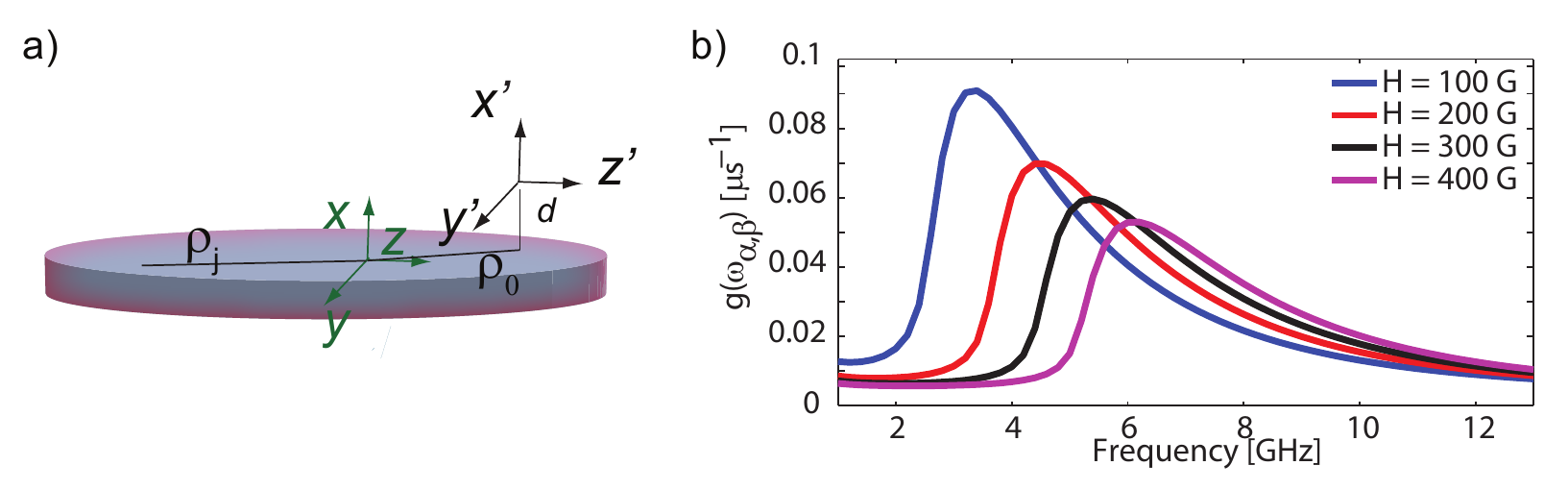}
\caption{ \textbf{a,} Reference frame $xyz$ for the disc and $x'y'z'$ for the stray-field probe. \textbf{b,} Field-dependence of the $g(\omega_{\alpha,\beta})$ rate for various magnetic fields, following eq. \eqref{eq:relratecomplicated} and using the parameters discussed in the text. A field-independent constant rate of $3.8(2) \cdot 10^{-3}$ $\mu s^{-1}$ was added to model the contribution of longitudinal spin fluctuations.}
\label{fig:FourMagn}
\end{figure}
\subsection{Stray-field detection of spin noise}
In this section, we will develop the model used to describe the relaxation data in Fig. 4 of the main text. We use the formalism of the previous section to derive a general expression for the stray magnetic-field noise generated by spin noise in a thin magnetic film.

Spin fluctuations $\delta \mathbf{S}_j(t)$ in the disc will produce a time-dependent field $\delta \mathbf{B'}(\mathbf{r}_{0},t)$ at the NV site, which can be written as:
\begin{align}
\delta \mathbf{B'	}(\mathbf{r}_{0},t) =  \mathscr{R}_{\theta_{\text{NV}}} \sum_j \mathscr{D}(\mathbf{r}_{j}-\mathbf{r}_0) \delta \mathbf{S}_j(t).
 \label{eq:BsumVar}
\end{align}
Here, we inserted a rotation matrix $\mathscr{R}_{\theta_{\text{NV}}}$ to express the field in an $x'y'z'$ frame which has the $z'$-axis parallel to the NV axis (i.e., the $x'y'z'$ frame is rotated around the $y$-axis with respect to the $xyz$ frame in \fref{FourMagn}a). We first compute the spectral density of the stray magnetic-field noise along a general direction $\eta$:
\begin{equation}
| B'_{{\eta},{\eta}}(\omega_{\alpha,\beta}) |^2  =  \int_{-\infty}^{\infty} \diff \tau e^{-i \omega_{\alpha,\beta} \tau} \overline{\delta B'_{\eta}(\tau) \delta B'_{\eta}(0)},
\label{eq:speNois}
\end{equation}
which has units of T$^2$/$\text{Hz}$. Here, $\overline{\langle \ldots \rangle}$ denotes an ensemble average over the magnet's spin degree of freedom. An expression for the stray magnetic-field noise can be obtained by inserting eq. \eqref{eq:BsumVar} into eq. \eqref{eq:speNois}: 
\be
| B'_{\eta,\eta}(\omega_{\alpha,\beta}) |^2 = \frac{\Gamma^2}{(2 \pi)^2} \int_{\mathbf{k}} \mathscr{N} (\mathbf{k},d) \cdot \mathbf{S}(\omega_{\alpha,\beta},\mathbf{k})  \diff \mathbf{k}. \label{eq:kernel1} 
\ee
where we have defined $\mathbf{S} = \{ S^{x,x}, S^{y,y}, S^{z,z} \}$, with 
\begin{equation}
S^{\eta_1,\eta_2}(\omega_{\alpha,\beta},\mathbf{r}_i,\mathbf{r}_j)  = \int_{-\infty}^{\infty} \diff \tau e^{-i \omega_{\alpha,\beta} \tau} \overline{\delta S^{\eta_1}_i(\tau)\delta S^{\eta_2}_j(0)}.
\label{eq:fluctuat}
\end{equation}
and
\be
\mathscr{N} (\mathbf{k},d) = \left( \begin{array}{ccc} 
\left| \mathscr{D}^{x,x}_{\mathbf{k}} \cos(\theta) - \mathscr{D}^{x,z}_{\mathbf{k}} \sin(\theta) \right|^2 &
\left| \mathscr{D}^{x,y}_{\mathbf{k}} \cos(\theta) - \mathscr{D}^{y,z}_{\mathbf{k}} \sin(\theta) \right|^2  & \left| \mathscr{D}^{x,z}_{\mathbf{k}} \cos(\theta) - \mathscr{D}^{z,z}_{\mathbf{k}} \sin(\theta) \right|^2 \\
\left| \mathscr{D}^{x,y}_{\mathbf{k}} \right|^2 & \left| \mathscr{D}^{y,y}_{\mathbf{k}} \right|^2 & \left| \mathscr{D}^{y,z}_{\mathbf{k}} \right|^2 \\
\left| \mathscr{D}^{x,z}_{\mathbf{k}} \cos(\theta) + \mathscr{D}^{x,x}_{\mathbf{k}} \sin(\theta) \right|^2 &
\left| \mathscr{D}^{y,z}_{\mathbf{k}} \cos(\theta) + \mathscr{D}^{x,y}_{\mathbf{k}} \sin(\theta) \right|^2  & \left| \mathscr{D}^{z,z}_{\mathbf{k}} \cos(\theta) + \mathscr{D}^{x,z}_{\mathbf{k}} \sin(\theta) \right|^2 \end{array} \right),
\label{eq:kernel2}
\ee
and the $\mathscr{D}^{m,n}_{\mathbf{k}} = \mathscr{D}^{m,n}(\mathbf{k},d)$ are the matrix elements of eq. \eqref{eq:dipoDefKernel}.

The matrix $\mathscr{N} (\mathbf{k},d)$ filters in $k$-space the spin fluctuations of the magnetic thin film. 
Note that the integral in eq. \eqref{eq:kernel1} contains a kernel $k^2 \exp(-2kd)$ for all the components of the $\mathscr{N} (\mathbf{k},d)$ matrix.
The detection annulus changes therefore slightly with respect to the case of static magnetometry. In the next section we discuss how the NV centre can be used to probe the spin noise via spin relaxation measurements.
\subsection{Noise probed by an NV centre}
\label{secA:relmatr}
In this section we first describe the fitting procedure used to extract the relaxation rates from the measurements in Fig. 4a of the main text. Then we describe the model linking the relaxation rates to the spin-noise in the disc.

To describe the NV-spin relaxation we use a rate-equation model\cite{Jarmola12}:
\begin{align}
\frac{\diff \mathbf{P}(t)}{\diff t} = \mathbf{\bar{W}}\mathbf{P}(t) =  \left( \begin{array}{ccc} -(W_{1,0}+W_{-1,0}) & W_{-1,0} & W_{1,0} \\
W_{-1,0} & -(W_{-1,0}+W_{1,-1}) & W_{1,-1} \\
W_{1,0} & W_{1,-1} & -(W_{1,-1}+W_{1,0})
\end{array} \right) \mathbf{P}(t)
\label{eq:master}
\end{align}
where $\mathbf{P}(t)$ describes the populations of the three NV-spin states as a function of time. In our fitting procedure, we imposed $W_{1,-1}=0$. This is validated by noting that magnetic-field noise does not directly couple the $m_s = -1,1$ levels. In addition, we observed that the relaxation of NV$_A$ is dominated by magnetic-field noise (it has a much faster relaxation than far-away NV$_{ref}$ which we confirmed in a separate measurement to be $\sim$1/ms).  The relaxation dynamics is therefore described by only two parameters: $W_{1,0}$ and $W_{-1,0}$.

In second order perturbation theory, the relaxation parameters are given by:\cite{Suter98}
\begin{equation}
W_{\alpha,\beta} \underset{\alpha \neq \beta}{=} \frac{1}{\hbar^2} \int_{-\infty}^{\infty} \diff \tau e^{i \omega_{\alpha,\beta} \tau} \overline{\langle \alpha| \mathscr{H}_1(\tau) | \beta \rangle \langle \beta | \mathscr{H}_1(0) | \alpha \rangle},
\label{eq:T1secondOrdr}
\end{equation}
where the $|\alpha\rangle$, $|\beta\rangle$ are the spin eigenstates of the NV centre and $\hbar \omega_{\alpha,\beta}$ is the energy difference between the levels ($\omega_{\alpha,\beta} = (\omega_{\alpha} - \omega_{\beta})$). 
The time-dependent magnetic perturbation at the NV centre due to magnetic-field fluctuations can be written as:
\begin{equation}
\mathscr{H}_1(\tau) = \hbar \gamma \sum_{\eta} \bar{I}_{\eta} \delta B_{\eta}(\tau),
\label{eq:coupli}
\end{equation}
where $\bar{I}_{\eta}$ is the $\eta$-component of the spin operator of the $I=1$ spin of the NV and $\gamma = 2\pi\cdot$28 GHz/T. It is easy to show that, by defining:
\begin{equation}
g(\omega_{\alpha,\beta}) = \frac{\gamma^2}{2} \int_{-\infty}^{\infty} \diff \tau e^{i \omega_{\alpha,\beta} \tau} \sum_{\eta \neq z}  \overline{\delta B_{\eta}(\tau) \delta B_{\eta}(0)} = \frac{\gamma^2}{2}  | B_{\perp}(\omega_{\alpha,\beta}) |^2,
\label{eq:spetralNoise}
\end{equation}
we have:
\begin{equation}
W_{1,0} = g(\omega_{+1}-\omega_{0})  \qquad W_{-1,0} = g(\omega_{-1}-\omega_{0}).
\end{equation}
Besides probing the spin-noise spectrum at different frequencies, the two relaxation channels are formally identical.

Combining \eref{kernel1} and \eref{spetralNoise}, we reach the following general expression for the characteristic relaxation rate of an NV centre in the vicinity of a thin magnetic film:
\begin{equation}
g(\omega_{\alpha,\beta}) = \frac{\gamma^2}{2} \frac{\Gamma^2}{(2 \pi)^2} \sum_{\substack{m \neq z\\n}} \int_{\mathbf{k}} \mathscr{N}^{m,n} (\mathbf{k},d) \cdot \mathbf{S}^{n}(\omega_{\alpha,\beta},\mathbf{k}) \diff \mathbf{k}.
\label{eq:finalResT1}
\end{equation}
Finally, note that the $T$- (temperature) and $B_{ext}$-dependent spin fluctuations ${S}^{\eta,\eta}(\omega_{\alpha,\beta},\mathbf{k})$ can be related to the more well-known spin susceptibility $\chi''_{\eta',\eta'}(\omega_{\alpha,\beta},\mathbf{k})$ with the fluctuation-dissipation theorem:\cite{Schwabl05}
\begin{align}
{S}^{\eta,\eta}(\omega_{\alpha,\beta},\mathbf{k}) = \frac{2 \hbar}{1 - e^{- \beta \hbar \omega_{\alpha,\beta}}} \chi''_{\eta',\eta'}(\omega_{\alpha,\beta},\mathbf{k}) \underset{\hbar \omega_{\alpha,\beta} \ll 1/\beta}{\approx}. \frac{2 k_{\mathrm{B}} T}{\omega_{\alpha,\beta}} \chi''_{\eta,\eta}(\omega_{\alpha,\beta},\mathbf{k})
\label{eq:fLDL}
\end{align}
\subsection{Noise due to a two-dimensional ferromagnetic thin film}
\noindent
In this section we evaluate eq. \eqref{eq:finalResT1} for the case of an infinitely extended 2d ferromagnetic layer of Permalloy. This procedure leads to the calculated noise spectrum and associated NV-relaxation rates shown in Fig. 4c of the main text (and to the corresponding solid lines in Fig. 4b of the main text.)

We will give an estimate of the characteristic relaxation time including at first only transverse spin fluctuations. 	
Assume the plane is magnetized along the $z'$-axis. The transverse susceptibility for a 2d ferromagnet is given by:\cite{Trifunovic13}
\begin{align}
\chi''_{y',y'}(\omega_{\alpha,\beta},k) = \Gamma^{-1} \sqrt{\frac{2}{\pi}} S \frac{W}{W^2 + (D k^2 + \Delta - h \omega_{\alpha,\beta})^2},
\label{eq:FMsusc}
\end{align}
where $D$ is the spin stiffness~\cite{Trifunovic13}, $W$ the width of the FMR excitation and $\Delta$ its energy. In the previous expression the $k$ in the denominator is expressed in reciprocal lattice units, namely it is adimensional, such that $D$ has units of energy.\\ 
When computing the integral in eq. \eqref{eq:finalResT1}, we are going to cast the $k^3 \diff k$ filter in dimensionful units by multiplying the result by $\Gamma^2$.\\
Because of the demagnetization energy cost for out-of-plane spin fluctuations, we can in addition safely assume that $\chi''_{y',y'} \gg \chi''_{x',x'}$ and include in the calculations only the contribution of $\chi''_{y',y'}$. \\
In the case of eq. \eqref{eq:FMsusc} the susceptibility has no $\phi_k$-dependence. We can therefore integrate out the $\phi_k$ variable in the integral \eqref{eq:finalResT1}. We define the following $\theta_{\text{NV}}$-dependent prefactor:
\begin{align}
F(\theta_{\text{NV}}) &= \left( \frac{\mu_0 \mu_{eff}}{2} \right)^2  \int_{0}^{2 \pi} \Big( \left| i \sin(\phi_k) \cos(\theta_{\text{NV}}) + \frac{\sin(2 \phi_k)}{2} \sin(\theta_{\text{NV}}) \right|^2 + \left| \sin^2(\phi_k) \right|^2 \Big) \diff \phi_k \nonumber \\
&= \frac{(\mu_0 \mu_{eff})^2}{32} \pi \Big( 11 + 3 \cos(2 \theta_{\text{NV}}) \Big).
\label{eq:tthetascal}
\end{align}
We can now solve the integral \eqref{eq:finalResT1}, which gives:
\begin{align}
g(\omega_{\alpha,\beta}) &= \frac{\gamma^2}{2}  \frac{\Gamma^3}{(2 \pi)^{2}} \left( \frac{2 k_{\mathrm{B}} T}{\omega_{\alpha,\beta}} \right)  S \sqrt{\frac{2}{\pi}}  F(\theta_{\text{NV}}) \int_0^{\infty} k^3 e^{-2 d k} \frac{W}{W^2 + (D k^2 + \Delta - h \omega_{\alpha,\beta})^2} \diff k \nonumber \\
&= \frac{\gamma^2}{2}  \frac{\Gamma^3}{(2 \pi)^{2}} \left( \frac{2 k_{\mathrm{B}} T}{\omega_{\alpha,\beta}} \right)  S \sqrt{\frac{2}{\pi}}  F(\theta_{\text{NV}})
\Big[ - \frac{W + i(\Delta - h \omega)}{4 D^2 \sqrt{\pi}} 
\MeijerG*{3}{1}{0}{0}{-1}{-1,0,1/2}{d^2\frac{-iW+\Delta -h \omega}{D}} \Big. \nonumber \\
\Big. &- \frac{W - i(\Delta - h \omega)}{4 D^2 \sqrt{\pi}} 
\MeijerG*{3}{1}{0}{0}{-1}{-1,0,1/2}{d^2\frac{iW+\Delta -h \omega}{D}} \Big],
\label{eq:relratecomplicated}
\end{align}
where the function $\MeijerG*{m}{n}{p}{q}{a_1, \dots, a_p}{b_1, \dots, b_q}{z}$ is the Meijer G-function and $d$ is expressed in lattice units. 
The previous expression formally holds true in the case of an infinitely large and thin magnetic film and it represents the contribution to the magnetic relaxation rate due solely to transverse spin fluctuations of the ferromagnet.\\
On the other hand, longitudinal spin fluctuations take into account thermal intra-band spin-wave transitions;\cite{Trifunovic13} due to the much smaller scale of the Zeeman with respect to the thermal energy, their contribution is essentially field-independent. We include them by summing up a constant to the expression \eqref{eq:relratecomplicated}.\\ 
For Permalloy we use the lattice constant $a = 3.55$ \AA\ (averaged between Ni and Fe). Accordingly, the effective moment can be written $\mu_{eff} = M_s a^3$, with $M_s = 8 \cdot 10^5$ A/m the saturation magnetization.\cite{Bayer2005} In addition, in micromagnetic simulations\cite{Bayer2005} the exchange stiffness is commonly set as $A = JS^2/a = 10^{-11}$ J/m. From $A$ and setting $S=1/2$, we obtain the spin stiffness\cite{Trifunovic13} $D = 2JS = 88$ meV. \\
For a comparison with the experimental data, we use $W = 0.2$ GHz, $d = 50$ nm and vary $\Delta(H)$ as the Kittel law measured in our experiments for a 6 $\mu$m large disc. For the comparison shown in the main text, we used the effective 2D spin density of $\Gamma = 5.6$ nm$^{-2}$ and fitted a field-independent constant relaxation rate of $3.8(2) \cdot 10^{-3}$ $\mu s^{-1}$. 
A plot of $g(\omega_{\alpha,\beta})$ with these parameters is in \fref{FourMagn}b. The full field-dependence of the spectrum $g(\omega_{\alpha,\beta})$ is shown in Fig. 4c of the main text. \\
\section{Numerical simulations}
In this section we describe the numerical calculations of the spatial magnetization profile used in this work. Fig. 1f of the main text presented such calculations for the \textit{static} magnetization of the disc. Fig. 2b of the main text shows the numerically calculated ferromagnetic resonance of the disc, and in Fig. 3 of the main text we presented numerical calculations of the time-averaged decrease in longitudinal magnetization of the disc.
\subsection{Static magnetization and conventions}
\noindent
Micromagnetic simulations have been performed with the open source software OOMMF\cite{OOMMF99} running on the Harvard Odyssey cluster. For all the results presented in this work, the magnetic properties of the Permalloy disc have been simulated imposing a spatial discretization of 5$\times$5$\times$30 nm$^3$. As in previous works,\cite{Davis10,Bailleul06} we chose a saturation magnetization for Permalloy of M$_s =$ 800 kA/m, an exchange coupling of A=10$^{-11}$ Jm$^{-1}$, a  Gilbert gyromagnetic ratio $\gamma = 2.21 \times 10^{5}$ m/As, a damping constant $\alpha = 0.005$. For the calculation of the static magnetic properties, the value of $\alpha$ has been increased to $\alpha = 0.95$ to favour convergence.\cite{OOMMF99}\\
Since the magnetic properties of a microdot are hysteretic~\cite{Davis10}, all measurements were done by first applying a large field and then sweeping the field down. To obtain Fig. 1e of the main text, we calculated the magnetization profile via the ODE solver of the Landau-Lifshitz equation. Then we calculated the associated disc stray field at the sites of the two NVs. For the calculation of the magnetization we imposed, at the largest computed field of 700 G, a field-aligned spin structure as initial configuration. Subsequently, the initial configuration at a lower field was chosen as the ground state of the adjacent field of bigger magnitude. In order to resemble experimental conditions, the external field orientation in the simulations has been chosen to be the same one of the nitrogen-vacancy center, e.g. along the $[\sin(\theta_{\text{NV}}),0,\cos(\theta_{\text{NV}})]$ direction, where the $z$-axis is the one normal to the magnet's surface and $\theta_{\text{NV}} = \arccos(1/\sqrt{3})$.

\subsection{Uniform dynamics}
\begin{figure}[h!]
\centering
\includegraphics[width=\textwidth]{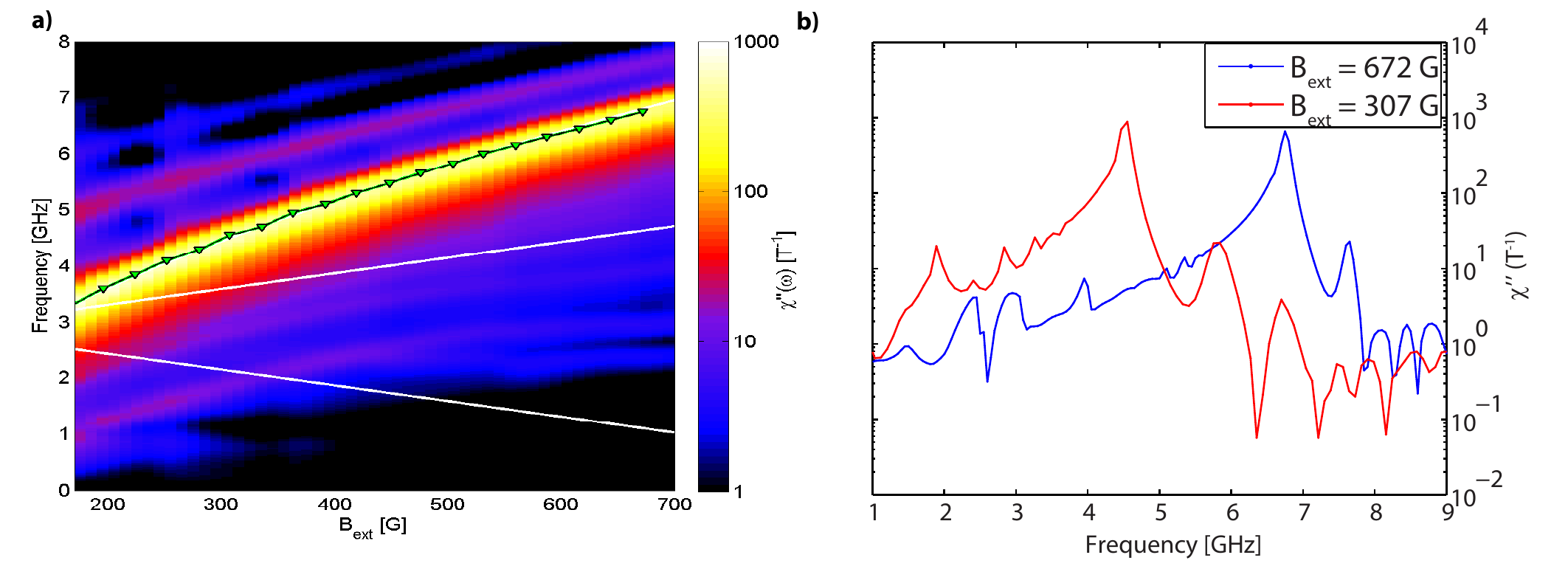}
\caption{\label{fig:colormapMode} \textbf{Linear transverse spin response of a Permalloy disc of 6$\mu$m in diameter a,} Colormap combining several $\chi_{\perp}''(\omega,B_{\text{ext}})$, where the value of $B_{\text{ext}}$ is the magnitude of the external field applied along the NV-axis direction. NV transitions are shown with white solid lines. Green triangles show the position of the uniform mode. \textbf{b,} Detail of $\chi_{\perp}''(\omega,B_{\text{ext}})$ at the fields $B_{\text{ext}} = 672$ and 307 G.}
\end{figure}

Here we describe how we calculate and define the ferromagnetic resonance (FMR) frequency of the disc as a function of the external magnetic field $B_{ext}$. We computed the linear, frequency-dependent, spatially-averaged, transverse magnetic susceptibility\cite{Schwabl05} $\chi_{\perp}''(\omega)$ by applying a spatially uniform magnetic pulse $h(t)$ of the duration $\Delta t = 50 \times 10^{-12}$ s oriented orthogonally to the plane of the disc (resembling the direction of the field generated by our coplanar waveguide). The following time-evolution $m_{\perp}(t)$ of the spatially integrated transverse disc's magnetization was then recorded for a total time of 20 ns, at $5 \times 10^{-12}$ s time intervals. We computed the field-dependent $\chi_{\perp}''(\omega)$ via a Fast Fourier transform $\mathscr{F}$ using the algorithm:
\begin{equation} 
\chi_{\perp}''(\omega) = \text{Im}\left(\frac{\mathscr{F}( m_{\perp}(t) - m_{\perp}(0) )}{\mathscr{F}(h(t))}\right),
\label{eq:Mperptot}
\end{equation}
with $h(t)$ expressed in Tesla and $m_{\perp}(t)$ in units of M$_s$. A colormap combining spectra at several fields is plotted in the left panel of Fig. \ref{fig:colormapMode}. Green triangles mark the position of the maximum spatially-averaged response, which represents the Kittel's law of the FMR for the disc.  Spectral power at two selected fields is shown in the right part of Fig. \ref{fig:colormapMode}. Well-isolated modes besides the uniform one can be recognized, along the lines of what was shown in previous studies of Permalloy micromagnets.\cite{Bailleul06} 
\subsection{Non-linear spin dynamics}
\begin{figure}[h!]
\centering
\includegraphics[width=\textwidth]{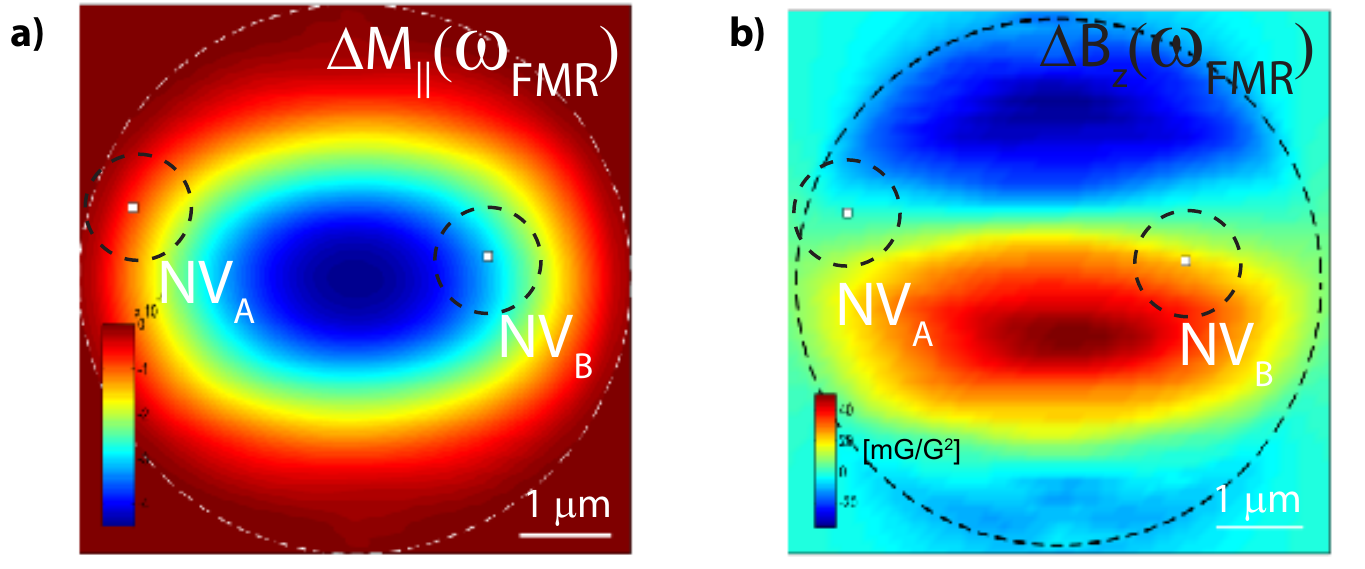}
\caption{\label{fig:ModeProfileBetter} \textbf{Time-averaged variation of the longitudinal magnetization and resulting stray field at the frequency $\omega_{\text{FMR}}$ of the uniform mode. a,} In-plane component $\Delta M_{\parallel}$ of the time-averaged quantity $\langle \delta m_{\parallel,i}(t) \rangle \mathbf{u}_{\parallel,i}$. Magnetization is normalized to M$_s$ units. NV transitions are also marked by white solid lines. \textbf{b,} Stray magnetic field along the NV-axis direction originating from the local magnetization $\langle \delta m_{\parallel,i}(t) \rangle \mathbf{u}_{\parallel,i}$. The stray field has been normalized by the Rabi one. The NV depth has been taken to be 50 nm.}
\end{figure}
\noindent
In order to understand the frequency dependence of the spin-wave magnetic field detected by the NV centre in Fig. 3 of the main text, it is necessary to quantitatively model the non-linear response of the ferromagnetic disc upon driving. We performed numerical simulations by imposing at each frequency $\omega$ a sinusoidal and out-plane continous excitation of the form $h(t) = h_0 \sin(\omega t)$. The static local magnetization $m_{\parallel,i}(t=\infty) \mathbf{u}_{\parallel,i}$ is therefore perturbed. \\
For the time evolution of $\mathbf{m}_{i}(t)$ we assumed the following ansatz:
\begin{equation}
\mathbf{m}_{i}(t) = \mathbf{u}_{\parallel,i} \left( m_{\parallel,i}(t=\infty) + \delta m_{\parallel,i}(t) \right) + \sum_{\eta} \mathbf{u}_{\eta,i} m_{\eta,i}(t),
\end{equation}
where $\eta$ labels the two axes orthogonal to $\mathbf{u}_{\parallel,i}$ and $\delta m_{\parallel,i}(t)$ is the non-linear change of the longitudinal magnetization. Provided with the previous assumption, we extracted from the numerical results the quantity $\delta m_{\parallel,i}(t)$ by using the algorithm:
\begin{equation}
\delta m_{\parallel,i}(t) = \left( \mathbf{m}_{i}(t) - \mathbf{m}_{i}(t=\infty) \right) \cdot \mathbf{u}_{\parallel,i}.
\end{equation}
The quantity $\delta m_{\parallel,i}(t)$ is then time-averaged in order to compute the stray field along the NV axis $\Delta B_z(\omega)$, to which our measurements are sensitive.
The quantity $\Delta B_z(\omega)$ is then casted into mG/G$^2$ by normalizing with the Rabi field $h_0 \sin(\theta_{\text{NV}})/\sqrt{2}$. In the left part of Fig.\ref{fig:ModeProfileBetter} we plot the exemplary in-plane component $\Delta M_{\parallel}$ of the time-averaged quantity $\langle \delta m_{\parallel,i}(t) \rangle \mathbf{u}_{\parallel,i}$ at the frequency $\omega_{\text{FMR}}$ of the uniform mode. The resulting stray field $\Delta B_z(\omega_{\text{FMR}})$ is shown in the right part of Fig.\ref{fig:ModeProfileBetter}. Similar plots were shown in Fig. 3c of the main text.

\end{document}